\documentclass[letterpaper,journal]{IEEEtran}
\usepackage{bm}
\usepackage[utf8]{inputenc}
\usepackage{graphicx}
\usepackage{epstopdf}
\usepackage{multirow}
\usepackage{amsmath}
\usepackage{amsfonts}
\usepackage{tabularx}
\usepackage{booktabs}
\usepackage{float}
\usepackage{array}
\usepackage{color}
\usepackage{amssymb}
\usepackage{cite}
\usepackage{soul}

\usepackage{adjustbox}
\usepackage{setspace}
\usepackage{makecell}
\usepackage{bbding}
\usepackage{utfsym}
\usepackage{epstopdf}
\usepackage{mathtools} 
\usepackage{mathtools}
\usepackage[ruled,linesnumbered,vlined,noend]{algorithm2e}

\usepackage{titlesec}
\usepackage{indentfirst}
\usepackage[caption=false,font=scriptsize,labelfont=sf,textfont=sf]{subfig}
\usepackage{booktabs} 
\usepackage{amsmath}  
\usepackage{cuted} 
\def\BibTeX{{\rm B\kern-.05em{\sc i\kern-.025em b}\kern-.08em
		T\kern-.1667em\lower.7ex\hbox{E}\kern-.125emX}}
\normalsize

\titlespacing*{\section}
  {0pt}                
  {0.5\baselineskip}     
  {0.5\baselineskip}   

\titlespacing*{\subsection}
  {0pt}                
  {0.5\baselineskip}   
  {0.5\baselineskip}

\titleformat{\subsubsection}[runin]
{\normalfont\itshape}{\arabic{subsubsection})}{0.5em}{}[\normalfont:~ ]

\ifCLASSINFOpdf
\else
\fi
\begin{document}

\thispagestyle{empty}
\pagestyle{empty}

%
\title{SIM-assisted Secure Mobile Communications via Enhanced Proximal Policy Optimization Algorithm}

\author{\IEEEauthorblockN{Wenxuan Ma,~\IEEEmembership{Student Member,~IEEE}, Bin Lin,~\IEEEmembership{Senior Member,~IEEE}, Hongyang Pan,~\IEEEmembership{Member,~IEEE}, \\Geng Sun,~\IEEEmembership{Senior Member,~IEEE}, Enyu Shi,~\IEEEmembership{Student Member,~IEEE}, Jiancheng An,~\IEEEmembership{Senior Member,~IEEE}, and 
Chau Yuen,~\IEEEmembership{Fellow,~IEEE}}

\thanks{This study is supported in part by the National Natural Science Foundation of China under Grant 62371085 and Grant 62471096, and in part by the Ministry of Education, Singapore, under its MOE Tier 2 under Grant T2EP50124-0032. (\textit{Corresponding author: Bin Lin}).\protect}

\thanks{Wenxuan Ma, Bin Lin and Hongyang Pan are with the Information Science and Technology College, Dalian Maritime University, Dalian 116026, China (e-mail: mawenxuan@dlmu.edu.cn; binlin@dlmu.edu.cn; panhongyang18@foxmail.com).  
\par Geng Sun is with the College of Computer Science and Technology, Jilin University, Changchun 130012, China, and also with the Key Laboratory of Symbolic Computation and Knowledge Engineering of Ministry of Education, Jilin University, Changchun 130012, China (e-mail: sungeng@jlu.edu.cn). 
\par Enyu Shi is with the School of Electronics and Information Engineering, Beijing Jiaotong University, Beijing 100044, China. (e-mail: enyushi@bjtu.edu.cn). \par Jiancheng An and Chau Yuen are with the School of Electrical and Electronics Engineering, Nanyang Technological University, Singapore 639798 (e-mail: jiancheng.an@ntu.edu.sg; chau.yuen@ntu.edu.sg).
\par This manuscript has been accepted by IEEE Transactions on Wireless Communications, DOI: 10.1109/TWC.2026.3658332
}}


\maketitle
\thispagestyle{empty}
\begin{abstract}
	With the development of sixth-generation (6G) wireless communication networks, the security challenges are becoming increasingly prominent, especially for mobile users (MUs). As a promising solution, physical layer security (PLS) technology leverages the inherent characteristics of wireless channels to provide security assurance. Particularly, stacked intelligent metasurface (SIM) directly manipulates electromagnetic waves through their multilayer structures, offering significant potential for enhancing PLS performance in an energy efficient manner. Thus, in this work, we investigate an SIM-assisted secure communication system for MUs under the threat of an eavesdropper, addressing practical challenges such as channel uncertainty in mobile environments, multiple MU interference, and residual hardware impairments. Consequently, we formulate a joint power and phase shift optimization problem (JPPSOP), aiming at maximizing the achievable secrecy rate (ASR) of all MUs. Given the non-convexity and dynamic nature of this optimization problem, we propose an enhanced proximal policy optimization algorithm with a bidirectional long short-term memory mechanism, an off-policy data utilization mechanism, and a policy feedback mechanism (PPO-BOP). Through these mechanisms, the proposed algorithm can effectively capture short-term channel fading and long-term MU mobility, improve sample utilization efficiency, and enhance exploration capabilities. Extensive simulation results demonstrate that PPO-BOP significantly outperforms benchmark strategies and other deep reinforcement learning algorithms in terms of ASR.
 
\end{abstract}

\IEEEpeerreviewmaketitle

%
%
\section{Introduction}

\par \IEEEPARstart{T}{he} sixth-generation (6G) wireless communication network is expected to support higher transmission rates, lower latency, and massive device connectivity, providing users with an unprecedented quality of service (QoS) \cite{10054381}. Such ubiquitous connectivity also introduces a new paradigm of user mobility, where users interact with the network in a much more dynamic and continuous manner. With the increasing number of mobile users (MUs), security issues for MUs in 6G scenarios are also becoming increasingly prominent \cite{DBLP:journals/ojcs/MucchiJCPSBMSAH21}, particularly the risk of malicious eavesdropping \cite{DBLP:journals/csm/ChortiBKZCSFP22}. Therefore, it is necessary to study secure communications to provide the desired QoS for MUs \cite{DBLP:journals/comsur/MaoLWK23}.

\par Due to the broadcast nature of wireless communications, they are vulnerable to interception by malicious eavesdroppers. As a result, traditional cryptographic methods such as symmetric key cryptography (SKC) and asymmetric key cryptography (AKC) solutions have been widely employed for enhancing data security \cite{DBLP:journals/tifs/Sun0LLNNWA23}. However, these classical cryptographic schemes are facing significant bottlenecks. For example, AKC is too computationally expensive for resource-constrained networks, while SKC requires pre-shared keys,  which is challenging in decentralized networks \cite{DBLP:journals/corr/abs-2507-06500}. Therefore, there is an urgent need to investigate the physical layer security (PLS), which avoids reliance on complex key management and ensures confidentiality by leveraging the inherent randomness, spatial variations and noise characteristics of the wireless communication channels \cite{DBLP:journals/tcom/GuoYZQZH19}. By doing so, PLS can effectively resist eavesdropping during signal transmission while significantly reducing the computational and communication overhead of the system \cite{DBLP:journals/comsur/NguyenLCHL21}. Nevertheless, the mobility of MUs introduces channel uncertainty, which further increases the challenges of PLS communications \cite{10536035}.

\par Reconfigurable intelligent surface (RIS) has recently emerged as a promising technology for future communication systems, offering the ability to shape intelligent and reconfigurable channel environments \cite{DBLP:journals/tifs/Sun0LLNNWA23, DBLP:journals/twc/ShuWWXYSSW24}. By adjusting the propagation of electromagnetic waves and actively reshaping wireless channels \cite{DBLP:journals/tmc/PanLSWGWNY25}, RIS can facilitate secure transmission \cite{DBLP:journals/tcom/ZhangCWLLS22}. For example, with passive beamforming, RIS can improve the secrecy rate by designing the reflected signals for legitimate users \cite{DBLP:journals/iotj/QinSWDGYS24}. However, RIS only focuses on a single-layer metasurface configuration, which significant limits the flexibility of achievable beamforming patterns \cite{DBLP:journals/tcom/YaoXXNYY23}, and further impacts the secrecy performance of users \cite{DBLP:journals/tifs/NiuLAZY25}.

\par Fortunately, stacked intelligent metasurfaces (SIMs) with multi-layer metasurfaces have garnered significant attention due to their ability to directly manipulate electromagnetic wave signals \cite{DBLP:journals/jsac/AnXNAHYH23}. By adjusting the phase of each meta-atom across different layers, SIMs enable highly efficient beamforming in wireless communication systems \cite{DBLP:journals/tcom/LiEXAYH25}. In PLS applications, an SIM can not only strengthen the desired signal for legitimate users through precise beamforming, but also suppress eavesdropping channels via electromagnetic-domain precoding \cite{DBLP:journals/corr/abs-2507-09575}. This capability reduces the computational and hardware complexity required for secure beamforming, providing an effective approach for constructing low-cost and low-power secure communication systems \cite{DBLP:journals/jsac/AnXNAHYH23, DBLP:journals/icl/NeriniC24}. Motivated by this observation, an SIM-based single-input single-output transmission scheme that enhances PLS while simultaneously improving transmission efficiency is investigated \cite{DBLP:journals/tifs/NiuLAZY25}. 

\par However, the multiple user interference will lead to intensified spectrum competition, thereby degrading communication security \cite{10153713}. Moreover, the mobility of MUs makes it difficult to obtain accurate channel state information (CSI), increasing the challenge of resisting eavesdropping \cite{10857476}. In addition, the practical residual hardware impairment (RHI) of MUs introduces complex non-linear interference rather than simple additive noise, which directly weakens signal power and exacerbates information leakage risks, further increasing the challenge of ensuring secure communications \cite{10516677}. To address the challenges mentioned above, an SIM-assisted secure communication system is investigated to maximize the achievable secrecy rate (ASR) for multiple MUs in the presence of an eavesdropper. The main contributions of this paper are summarized as follows:

\begin{itemize}
        \item \textbf{\textit{SIM-assisted Secure Communication System for MUs}}: We investigate an SIM-assisted uplink secure communication system with multiple MUs, where the SIM is integrated into the base station (BS) to receive signals from MUs with time-varying positions. Meanwhile, a stationary eavesdropper attempts to intercept the information directly from the MUs. To enhance the practicality of the system, both inter-MU interference and RHI are also taken into account. To the best of our knowledge, it is the first work to propose an SIM-assisted system that improves security performance by real-time adjustment of phase shift and transmit power, enabling dynamic response to the time-varying positions of user mobility and potential threats.

    	\item \textbf{\textit{Problem Formulation to Maximize ASR through Jointly Adjusting Power and Phase Shift}}: Owing to the mutually coupled variables in the considered system, we formulate a joint power and phase shift optimization problem (JPPSOP) to maximize the average ASR for all MUs across the entire communication period, satisfy to practical constraints of each MU minimum ASR requirement and the maximum transmit power limits. The formulated problem explicitly considers the sequential and interdependent nature of decisions across multiple time slots, which is a critical consideration in a dynamic mobile environment. Thus, the long-term nature of the problem necessitates a careful trade-off between short-term and sustainable performance gains over the entire operational period.

       \item \textbf{\textit{Improved Deep Reinforcement Learning Approach}}: Due to the limitations of the conventional algorithm in handling user mobility, sample efficiency, and evaluation bias, we introduce the bidirectional long short-term memory (Bi-LSTM) mechanism to capture historical mobility trends, the off-policy data utilization (OPDU) mechanism to accelerate adaptation to time-varying channels, and the policy feedback (PF) mechanism to ensure consistent ASR optimization, respectively. Specifically, we propose a proximal policy optimization method with a Bi-LSTM mechanism, an OPDU mechanism, and a PF mechanism (PPO-BOP) to solve JPPSOP. Extensive simulation results validate the effectiveness of the proposed PPO-BOP.
\end{itemize}
\par The remainder of this paper is organized as follows: Section \ref{Related work} reviews related work. Section \ref{System model} presents the system model and formulates the problem. Section \ref{DRL-Based Approach} proposes the PPO-BOP. Section \ref{Simulation results and analysis} provides the simulation results. Finally, Section \ref{Conclusion} concludes the paper.
\par \textit{Notations:} The scalars are denoted by lowercase italic letters $x$, column vectors by boldface lowercase letters $\mathbf{x}$, and matrices by boldface uppercase letters $\mathbf{X}$. The superscripts $(\cdot)^{\mathrm{T}}$ and $(\cdot)^{\mathrm{H}}$ denote the transpose and conjugate transpose, respectively. $\mathbb{C}^{M \times N}$ represents the space of $M \times N$ complex matrices. $\mathbb{E}\{\cdot\}$ and $\nabla$ denote the statistical expectation and gradient operators, respectively. $\|\cdot\|_2$ represents the Euclidean norm, $|\cdot|$ denotes the absolute value of a scalar or the cardinality of a set, and $\text{diag}(\cdot)$ indicates a diagonal matrix. The $\mathrm{mod}$ is a modulo operation. The notation $x \sim \mathcal{CN}(\mu, \sigma^2)$ signifies that the random variable $x$ follows a circularly symmetric complex Gaussian distribution with mean $\mu$ and variance $\sigma^2$. Furthermore, $\lfloor \cdot \rfloor$ denotes the floor function, $[x]^+ \triangleq \max\{x, 0\}$, and $j \triangleq \sqrt{-1}$ is the imaginary unit.
\section{Related work}
\label{Related work}
\subsection{RIS-assisted Secure Communications} 
\par There were some works considering RIS and its evolved versions to enhance PLS in static communication scenarios \cite{DBLP:journals/tifs/ZhangDSAN21}. For example, Ge \emph{et al}. \cite{DBLP:journals/tvt/GeF23} investigated secure communications in ground-based cognitive satellite networks and proposed a novel robust cooperative beamforming scheme assisted by an active RIS. Salem \emph{et al}. \cite{DBLP:journals/tvt/SalemII23} examined the secure operation of an active RIS-assisted integrated sensing and communication (ISAC) system. Meanwhile, Tang \emph{et al}. \cite{DBLP:journals/twc/TangCLDZCDW24} focused on simultaneous transmission and reflection of reconfigurable intelligent surface (STAR-RIS)-assisted wireless communication systems and proposed a novel energy splitting transmission scheme, integrating it with uplink non-orthogonal multiple access (NOMA) technology. Furthermore, Zhang \emph{et al}. \cite{10057422} investigated robust transmission schemes for an RIS-assisted NOMA secure network in the presence of transceiver hardware impairments and imperfect successive interference cancellation. Salem \emph{et al}. \cite{DBLP:journals/ojcs/SalemII24} proposed a strategy for secure transmission in RIS-assisted full duplex cognitive radio systems by coordinating beamforming and power control. Similarly, in \cite{DBLP:journals/tcom/Lv0H023}, the authors investigated an RIS-assisted uplink NOMA machine-type communication network operating in the millimeter wave frequency band, with particular emphasis on secure transmission under finite block-length conditions. Shu \emph{et al}. \cite{DBLP:journals/twc/ShuWWXYSSW24} proposed an RIS-assisted hybrid secure spatial modulation system, which aims to maximize the system secrecy rate by jointly optimizing the RIS beamforming and the hybrid precoding of the transmitter. Zhang \emph{et al}. \cite{DBLP:journals/tcom/ZhangCWLLS22} studied the security enhancement of an RIS-assisted NOMA network by maximizing the minimum secrecy rate of the users, where the probability of secrecy outage was adopted as a security metric and instantaneous CSI of the eavesdropper was considered.
\par In contrast to these static settings, securing communications for MUs has also attracted growing attention in RIS-aided networks. Wan \textit{et al}. \cite{DBLP:journals/tifs/WanLCHWCYJ24} exploited the full space coverage of STAR-RIS to enable physical layer key generation for MUs, effectively resolving the coverage blind spots inherent in traditional RIS. Building on the advantages of this architecture, focusing on internal eavesdropping scenarios where MUs acted as potential eavesdroppers against each other, Shang \textit{et al}. \cite{10735389} investigated an STAR-RIS-aided secure communication framework. Zheng \textit{et al}. \cite{DBLP:journals/twc/ZhengMTYLW24} proposed a co-design of transmit diversity and active precoding for RIS-aided systems, ensuring reliable communication for MUs without requiring CSI. In terms of resource management, Hashida \textit{et al}. \cite{9681871} proposed a mobility aware user association strategy that used trajectory prediction to balance capacity and reliability for MUs. Zhang \textit{et al}. \cite{10970034} minimized the energy consumption of an RIS-assisted uncrewed aerial vehicle (UAV)-mobile edge computing (MEC) network while guaranteeing secure task offloading for MUs. Similarly, Guo \textit{et al}. \cite{10376206} enhanced the security of uplink NOMA communications for MUs by employing a UAV-mounted STAR-RIS and optimizing its flight trajectory. Qin \emph{et al}. \cite{DBLP:journals/iotj/QinSWDGYS24} investigated the security problem of an STAR-RIS-assisted NOMA uplink system in the presence of a cooperative jammer and two eavesdroppers. To maximize the sum secrecy rate, they proposed a deep reinforcement learning (DRL) algorithm.
\par However, despite these advancements, RIS exhibits certain limitations in improving secure communications. In mobile scenarios, the limited beamforming flexibility of single-layer metasurfaces hinders inter-MU interference mitigation, which in turn introduces significant complexity to the enhancement of legitimate MU signals. Consequently, it is necessary to explore more advanced multi-layer architectures for complex secure mobile communication.
\subsection{SIM-assisted Wireless Communications}
\par Distinct from the previously discussed RIS and STAR-RIS, SIM is capable of performing advanced signal processing directly in the electromagnetic wave domain, eliminating the need for digital beamformers and significantly improving the resolution and flexibility of beamforming \cite{DBLP:journals/jsac/AnXNAHYH23, DBLP:journals/wcl/LinAGDY24}. Niu \emph{et al}. \cite{DBLP:journals/tifs/NiuLAZY25} proposed deploying SIM at the end of signal transmission to introduce additional spatial degrees of freedom into the single-input single-output system, thus enabling secure communication without the need for costly radio frequency chains. This study demonstrated the effectiveness of SIM in secure communications. To further extend the application of SIM to broader domains, Hu \emph{et al}. \cite{DBLP:journals/tvt/HuZSLAYA25} investigated its role in the cell-free massive multiple-input multiple-output (CF-mMIMO) scenario. They conducted an in-depth study on joint beamforming and power allocation in CF-mMIMO systems, aiming to maximize overall system throughput. Niu \emph{et al}. \cite{DBLP:journals/wcl/NiuAPGCD24} extended SIM application to the ISAC domain. They investigated SIM-assisted ISAC systems and demonstrated that SIM could achieve transmit precoding in the wave domain. Jiang \emph{et al}. \cite{11271837} introduced an SIM architecture for satellite based ISAC to enable efficient wave domain beamforming. The study used a gradient ascent algorithm to maximize communication rates under sensing constraints. Zhang \emph{et al}. \cite{11079693} introduced an SIM-enabled uplink transmission framework to enhance reliability and latency performance for internet of things devices operating in the finite blocklength regime.
\par Beyond these system-level integrations, other studies have investigated the use of SIM for specific signal processing tasks. In \cite{DBLP:journals/jsac/AnYGRDPH24}, the authors proposed a novel method for the estimation of the two-dimensional direction of arrival using an SIM. By designing the phase shift of the SIM input layer, the system could automatically compute the discrete two-dimensional Fourier transform during the propagation of incident waves. Jia \emph{et al}. \cite{DBLP:journals/corr/abs-2502-05819} applied SIM to a multiple user beam-focusing scenario in the near field communications. They adopted a more practical spherical wavefront model to characterize near-field propagation and proposed replacing the conventional digital baseband architecture with an SIM. 
With a broad understanding of the fundamental theories of SIM, the research focus had gradually shifted to the optimization of its structural designs and the exploration of cutting-edge application scenarios. Darsena \emph{et al}. \cite{DBLP:journals/ojcs/DarsenaVIG25} proposed a novel SIM-assisted multiple user downlink transmission scheme, which ingeniously integrated a near-passive phase control layer with an active amplitude control layer incorporating an embedded amplifier chip within the SIM. Additionally, Nerini \emph{et al}.\cite{DBLP:journals/icl/NeriniC24} proposed a physically consistent SIM channel model with mutual coupling, showing that a single layer beyond the diagonal RIS could reach the performance upper bound. Xiong \emph{et al}. \cite{DBLP:journals/tnse/XiongCXQLY26} proposed a digital twin-based framework integrating SIM technology and composite potential fields for advanced air mobility. This method jointly optimized beamforming configurations and electric air taxi trajectories using deep Q-learning to enhanced ground-to-air transmission rates and aviation safety. Li \emph{et al}. \cite{11276870} proposed a fully-analog SIM architecture to perform wave-domain beamforming for wideband MIMO systems. Phase shifts were optimized by a block coordinate descent and penalty convex concave procedure algorithm to minimize channel fitting errors.
\par However, all of the abovementioned works ignored the secure mobile communications, where the stacked architecture and reconfigurable wavefront manipulation of SIMs can provide additional degrees of freedom for enhancing both communication quality and PLS. This gap highlights the necessity of investigating SIM-assisted secure communication schemes in dynamic mobile scenarios, particularly under varying user mobility and channel conditions.


\section{System Model and Problem Formulation}
\label{System model}

\par
\begin{figure}[!t]
\centering
\includegraphics[width=3.4in]{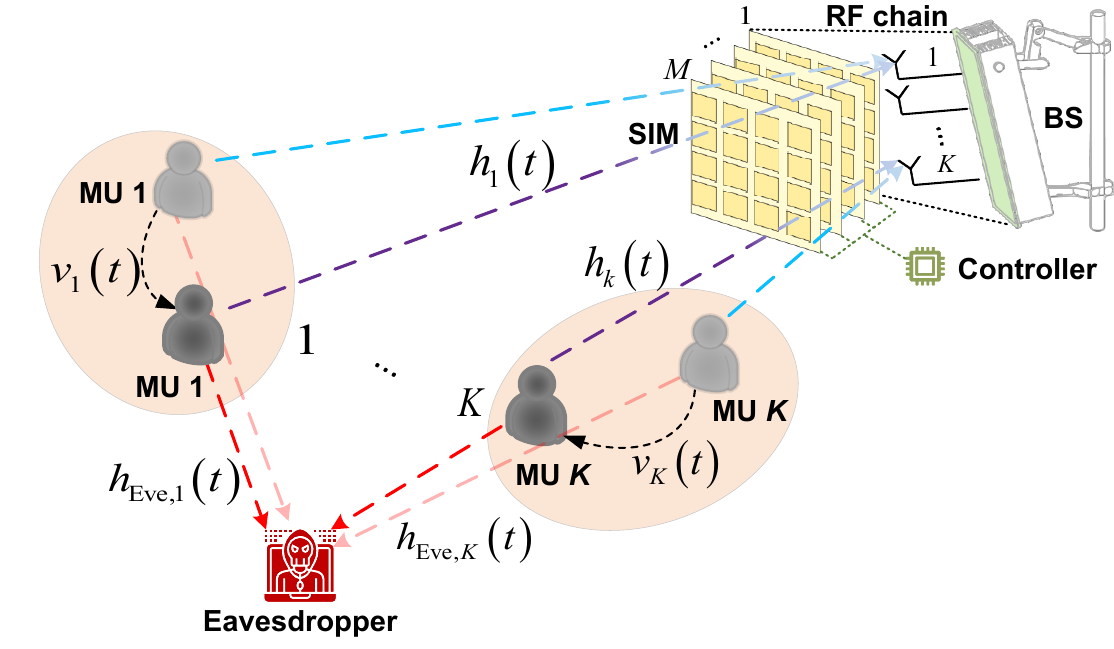}
\caption{An SIM-assisted secure communication system for MUs.}
\label{fig1}
\end{figure}

 As shown in Fig. \ref{fig1}, an SIM-assisted secure communication system is investigated, where an SIM is integrated into the radome of a BS to receive uplink data from $K$ MUs in the presence of an eavesdropper. 
An intelligent controller in the BS enables the application of distinct and adjustable phase shift to the electromagnetic waves traversing each meta-atom \cite{shi2025downlink}. Moreover, we assume that the MUs and the eavesdropper are equipped with a single antenna, while the BS is equipped with multiple antennas. The SIM is composed of $M$ metasurface layers, each layer containing $N$ meta-atoms. Let $\mathcal{K}=\{1,...,K\}$, $\mathcal{M}=\{1,...,M\}$, and $\mathcal{N}=\{1,...,N\}$ denote the index sets of the MUs and the BS antennas, the SIM metasurface layers, and the meta-atoms within each layer, respectively. To describe the mobility of MUs, we assume the total duration $L$ is divided into $T$ time slots, with the duration of each time slot being $\Delta t = L/T$. We assume $\Delta t$ is small enough such that the positions of the MU and the eavesdropper remain approximately constant at each time step $t$. The velocity of MU $k$ is denoted by $v_k(t)$, and its three-dimensional (3D) coordinates are given by $\mathbf{q}_k(t) = [X_{k}(t), Y_{k}(t),0]$. The eavesdropper and the BS are denoted as $\mathbf{q}_\mathrm{Eve} = [X_\text{Eve}, Y_\text{Eve},0]$ and $\mathbf{q}_\mathrm{BS} =(X_\mathrm{BS}, Y_\mathrm{BS}, Z_\mathrm{BS})$, respectively. 

\subsection{SIM Model}
\label{SIM Model}
\par We use $e^{j\varphi_{m}^{n}(t)}$ ($\forall m \in \mathcal{M}$, $\forall n \in \mathcal{N}$, $\varphi_{m}^{n}(t) \in [0, 2\pi)$) to denote the phase shift induced by the $n$-th meta-atom of the $m$-th metasurface layer of the SIM within the $t$-th time slot \cite{DBLP:journals/jsac/AnXNAHYH23}. Then, the diagonal phase shift $\mathbf{\Phi}_m(t) \in \mathbb{C}^{N \times N} $ for the $m$-th metasurface layer at the SIM can be represented as follows:
\begin{equation} 
\label{eq:1}
\begin{split}
    \mathbf{\Phi}_m(t) = \text{diag}(e^{j\varphi_{m}^{1}(t)}, e^{j\varphi_{m}^{2}(t)}, \dots, e^{j\varphi_{m}^{N}(t)}).
\end{split}
\end{equation}
\par Furthermore, $\mathbf{w}^{1}_{k} \in \mathbb{C}^{N\times 1 }$ denotes the transmission vector from the $k$-th antenna of the BS to the first metasurface layer within the SIM. Let $\mathbf{W}_m \in \mathbb{C}^{N \times N}, \forall m \in \mathcal{M}, \forall m \neq 1$ denote the transmission matrix from the $(m-1)$-th metasurface layer to the $m$-th metasurface layer of the SIM \cite{DBLP:journals/corr/abs-2507-09575}, in which  the element in the $n'$-th row and $n$-th column of $\mathbf{W}_m$ represents the corresponding diffraction coefficient between the $n'$-th meta-atom on the $(m-1)$-th layer and the $n$-th meta-atom on the $m$-th layer, denoted as $w_m^{n,n'}$, which is expressed as follows \cite{lin2018all}:
\begin{equation}
\label{eq:2}
\begin{split}
    w_m^{n,n'} = \frac{d_x d_y \cos \chi_m^{n,n'}}{d_m^{n,n'}} \left( \frac{1}{2\pi d_m^{n,n'}} - j \frac{1}{\lambda} \right) e^{j 2\pi \frac{d_m^{n,n'}}{\lambda}},
\end{split}
\end{equation}
\noindent where $\lambda$ represents the carrier wavelength, $d_m^{n,n'}$ denotes the corresponding transmission distance, $d_x \times d_y$ represents the dimensions of each SIM meta-atom, and $\chi_m^{n,n'}$ denotes the angle between the propagation direction and the normal direction of the $(m-1)$-th SIM metasurface layer \cite{DBLP:journals/wcl/LinAGDY24}. Similarly, the $n$-th element $w_{n,1}^1$ of $\mathbf{w}_k^1$ can be obtained from Eq. (\ref{eq:2}) \cite{shi2025downlink}. Then, the wave-based beamforming matrix $\mathbf{G}(t) \in \mathbb{C}^{N \times N}$ realized by the SIM at the BS can be obtained as:
\begin{equation} 
\label{eq:3}
\begin{split}
    \mathbf{G}(t) = \mathbf{\Phi}_M(t) \mathbf{W}_M \mathbf{\Phi}_{M-1}(t) \mathbf{W}_{M-1} \cdots \mathbf{\Phi}_2(t) \mathbf{W}_2 \mathbf{\Phi}_1(t).
\end{split}
\end{equation}

\subsection{Channel Model}
\label{Channel Model}

\par We consider a quasi-static flat-fading channel model. Let $\mathbf{h}_{\mathrm{SIM}, k} \in \mathbb{C}^{N \times 1}, \forall k \in \mathcal{K}$ denote the direct channel from the MU $k$ to the last metasurface layer of the SIM, which is assumed to follow a spatially correlated Rician fading channel model \cite{shi2025downlink}:
\begin{equation}
\label{eq:sim_k}
    \mathbf{h}_{\mathrm{SIM}, k}(t) = \sqrt{\frac{\beta_{\mathrm{SIM}, k}(t)}{1 + \xi_{\mathrm{SIM}, k}}} \left(\sqrt{\xi_{\mathrm{SIM}, k}} \bar{\mathbf{h}}_{\mathrm{SIM}, k}(t) + \tilde{\mathbf{h}}_{\mathrm{SIM}, k}(t)\right),
\end{equation}
\noindent where $\beta_{\mathrm{SIM}, k}$ is the path loss coefficient. $\xi_{\mathrm{SIM}, k}$ is the Rician factor, representing the power ratio between the line-of-sight (LoS) component and the non-LoS (NLoS) component of the corresponding path. $\bar{\mathbf{h}}_{\mathrm{SIM}, k}(t) \in \mathbb{C}^{N \times 1}$ represents the LoS component. $\tilde{\mathbf{h}}_{\mathrm{SIM}, k}(t) \sim \mathcal{CN}(\mathbf{0}, \mathbf{R}) \in \mathbb{C}^{N \times 1}$ is the NLoS component, where $\mathbf{R} \in \mathbb{C}^{N \times N}$ represents the spatial correlation among different meta-atoms of the output metasurface layer within the SIM \cite{9300189}. Without loss of generality, we assume an isotropic scattering environment where the multipath components are uniformly distributed. Therefore, the $(n, n')$-th element of $\mathbf{R}$ is given by $\mathbf{R}_{n, n^{\prime}} = \operatorname{sinc}(2 d_{n, n^{\prime}} / \lambda)$, where $d_{n, n^{\prime}}$ denotes the distance between meta-atoms, and $\operatorname{sinc}(x) = \sin (\pi x) / (\pi x)$ denotes the normalized sinc function. For the LoS channel, it is modeled as follows \cite{DBLP:journals/wcl/NiuAPGCD24, DBLP:journals/wcl/PapazafeiropoulosKKV25}:
\begin{equation}
\label{eq:los_model}
\begin{split}
    \bar{\mathbf{h}}_{\mathrm{SIM}, k}(t) = \mathbf{a}_{N}(\psi_{k}^{a}(t), \psi_{k}^{e}(t)),
\end{split}
\end{equation}
\noindent where $\psi_{k}^{a}$ and $\psi_{k}^{e}$ are the elevation and azimuth angles of arrival (AoAs) of the incident signal from MU $k$ to the SIM, respectively, and $n$-th element is given by \cite{DBLP:journals/wcl/PapazafeiropoulosKKV25}:
\begin{equation}
\label{eq:a_N_phi}
\begin{aligned}
    [\mathbf{a}_{N}(\psi_{k}^{a}(t), \psi_{k}^{e}(t))]_n = {} & \exp \bigg\{ j 2\pi \frac{d_{\mathrm{SIM}}}{\lambda}
    \bigg( \left\lfloor \frac{(n-1)}{\sqrt{N}} \right\rfloor\\ \sin \psi_k^{e}(t) \sin \psi_k^{a}(t) &+ \left((n-1) \bmod \sqrt{N} \right) \cos \psi_k^{e}(t) \bigg)
    \bigg\},
\end{aligned}
\end{equation}
\noindent where $d_{\mathrm{SIM}}$ is the spacing between the adjacent elements on each metasurface of the SIM \cite{ DBLP:journals/corr/abs-2408-04837}. Then, the overall channel between the BS and the MU $k$, $h_{k}(t)$ can be expressed as
\begin{equation}
\label{eq:overall_channel_k}
\begin{split}
    h_{k}(t) = \mathbf{w}_k^{1\mathrm{H}}
     \mathbf{G}(t)^\mathrm{H} \mathbf{h}_{\mathrm{SIM}, k}(t).
\end{split}
\end{equation}
\par Let ${h}_{\mathrm{Eve}, k}(t)$ denote the direct channel from the MU to the eavesdropper, which can be expressed as follows:
\begin{equation}
\label{eq:Eve_k}
    h_{\mathrm{Eve}, k}(t) = \sqrt{\frac{\beta_{\mathrm{Eve}, k}(t)} {1 + \xi_{\mathrm{Eve}, k}}} \left(\sqrt{\xi_{\mathrm{Eve}, k}} \bar{h}_{\mathrm{Eve}, k}(t) + \tilde{h}_{\mathrm{Eve}, k}(t)\right),
\end{equation}
\noindent where $\bar{h}_{\mathrm{Eve}, k}(t)$ represents the LoS component. $\tilde{h}_{\mathrm{Eve}, k}(t) \sim \mathcal{CN}(\mathrm{0}, \mathrm{1})$ represents the NLoS component, defined as the independent modeling of the random scattering component by complex Gaussian circularly symmetric random variables with a unit variance and a zero mean for all links \cite{DBLP:journals/tvt/HuZSLAYA25}. 
\par Additionally, the loss of the path of the channel from MU $k$ to BS is $\beta_{\mathrm{SIM}, k}(t) = \rho d_{\mathrm{SIM},k}^{-\alpha_{\mathrm{SIM},k}}(t)= \rho ||\mathbf{q}_k(t) - \mathbf{q}_\mathrm{BS}||_2^{-\alpha_{\mathrm{SIM},k}}$, where $\rho$ is the constant path loss at the reference distance, $\alpha_{\mathrm{SIM},k}$ is the path loss exponent. $d_{\mathrm{SIM},k}(t)$ is the distance from the MU to the BS. Furthermore, the loss of the path of the channel from the MU to the eavesdropper is $ \beta_{\mathrm{Eve}, k}(t) =\rho d_{\mathrm{Eve},k}^{-\alpha_{\mathrm{Eve},k}}(t)= \rho ||\mathbf{q}_k(t) - \mathbf{q}_\mathrm{Eve}||_2^{-\alpha_{\mathrm{Eve},k}}$,
where $d_{\mathrm{Eve},k}(t)$ is the distance from the MU to the eavesdropper, $\alpha_{\mathrm{Eve},k}$ is the path loss exponent \cite{DBLP:journals/tmc/ZhangSWLLNL24}.
\par Then, the signal transmitted by MU $k$ is represented as
\begin{equation}
\label{eq11}
\begin{split}
    x_k(t) = \sqrt{P_k(t)} s_k(t) + \eta_k(t),
\end{split}
\end{equation}
\noindent where $\eta_k(t)$ is the aggregate distortion noise caused by the non-ideal transmitter, which can be modeled as
$ \eta_k(t) \sim \mathcal{CN}(0, \kappa_k^2 P_k) $. This aggregate distortion stems from practical transceiver imperfections such as power amplifier non-linearity, phase noise, and in-phase/quadrature imbalance \cite{DBLP:journals/comsur/MaoLWK23}, \cite{DBLP:journals/tifs/GuoJCW24}. Furthermore, $\kappa_k$ represents the level of RHI, which can be measured of the error vector. Then, the signal $y_k(t)$ received at the BS can be expressed as
\begin{equation}
\label{eq12}
\begin{split}
    y_k(t) &= \sum_{k=1}^{K} h_k(t)x_k(t) + \tilde{n}_b(t) \\
    &= \mathbf{w}_k^{1\mathrm{H}}\sum_{k=1}^{K}  \mathbf{G}(t)^\mathrm{H} \mathbf{h}_{\text{SIM},k}(t) x_k(t) + \tilde{n}_b(t).
\end{split}
\end{equation}
\par Similarly, the signal received by the eavesdropper is expressed by
\begin{equation}
\label{eq13}
\begin{split}
    y_{\mathrm{Eve},k}(t) &= h_{\mathrm{Eve}, k}(t)x_k(t) + \hat{n}_b(t),
\end{split}
\end{equation}
\noindent where $\tilde{n}_b(t)$ and $\hat{n}_b(t)$ represents additive white Gaussian noise (AWGN) with zero mean and variance $N_0$, $ \tilde{n}_b(t) \sim \mathcal{CN}(0, N_0) $, $ \hat{n}_b(t) \sim \mathcal{CN}(0, N_0) $.

\subsection{User Mobility Model}
\label{Users Mobility Model}
\par To better model the mobility of real-world MUs, we adopt the random walk-based model to simulate the mobility of the MU, with a bias introduced in the direction update. Specifically, within each time slot $\Delta t$, MUs randomly update their direction of movement $\theta_k(t)$ and velocity $v_k(t)$ according to predefined parameters, thereby ensuring that the trajectory of the MUs follows a persistent directional trend \cite{DBLP:journals/comsur/TabassumSH19}. Specifically, the update rule for $\theta_k(t)$ is as follows:
\begin{equation}
\label{eq14}
\begin{split}
    \theta_k(t) = \theta_k^{\text{fix}} + \Delta\theta_k(t),
\end{split}
\end{equation}
\noindent where $\theta_k^{\text{fix}}$ is the fixed reference direction angle\cite{DBLP:journals/jsac/HashidaKKIM22}. $\Delta\theta_k(t) $ is a random angular perturbation at the beginning of $\Delta t$, uniformly distributed in the interval $[-\Delta\Theta, \Delta\Theta]$. 
\par Likewise, $v_k(t) \in [0, V_{\max}]$ is randomly updated in $\Delta t$. Then, the displacement components for the MU moving along $\theta_k(t) $ within $\Delta t$ can be expressed as follows:
\begin{equation}
\label{eq:15}
\begin{split}
    \Delta X_k(t) = v_k(t)\Delta t \cos(\theta_k(t)), \\ 
    \Delta Y_k(t) = v_k(t)\Delta t \sin(\theta_k(t)),
\end{split}
\end{equation}
\noindent through the displacement components, and the position coordinates of the MU can be further expressed as follows: 
\begin{equation}
\label{eq16}
\begin{split}
    X_k(t+1) = X_k(t) + \Delta X_k(t), \\
    Y_k(t+1) = Y_k(t) + \Delta Y_k(t).
\end{split}
\end{equation}

\subsection{Uplink Secrecy Rate Model}
\label{Uplink Secrecy Rate}

\par We can obtain the complete signal received in BS by substituting Eq. (\ref{eq11}) in Eq. (\ref{eq12}), which is given by
\begin{equation}
\label{eq17}
\begin{alignedat}{1}
&y_k(t)\\
&= \mathbf{w}_k^{1\mathrm{H}}\sum_{k=1}^{K}
  \mathbf{G}(t)^\mathrm{H}\mathbf{h}_{\mathrm{SIM},k}(t)
  \bigl(\sqrt{P_k(t)}\,s_k(t)+\eta_k(t)\bigr)
  + \tilde{n}_b(t)\\
&= \mathbf{w}_k^{1\mathrm{H}}\sum_{k=1}^{K}
  \mathbf{G}(t)^\mathrm{H}\mathbf{h}_{\mathrm{SIM},k}(t)
  \sqrt{P_k(t)}\,s_k(t)\\
&\quad
  + \mathbf{w}_k^{1\mathrm{H}}\sum_{k=1}^{K}
  \mathbf{G}(t)^\mathrm{H}\mathbf{h}_{\mathrm{SIM},k}(t)
  \eta_k(t)
  + \tilde{n}_b(t)\,.
\end{alignedat}
\end{equation}

\par The signal-to-interference-plus-noise ratio (SINR) at BS and the SINR at the eavesdropper from the signal of MU $K$ are given in Eq. (\ref{eq18}) and Eq. (\ref{eq19}), respectively, as shown at the top of the next page, where $\sum_{i=1}^{K} \left| \mathbf{w}_i^{1\mathrm{H}} \mathbf{G}(t) \mathbf{h}_{\text{SIM},i}(t) \right|^2 \kappa_i^2 P_i(t) $ represents the interference power caused by RHI \cite{DBLP:journals/tifs/GuoJCW24}. Then, the data rate for the $k$-th MU is
\addtocounter{equation}{2} 
\begin{equation}
\label{eq20}
\begin{split}
    R_k(t) = \log_2(1 + \gamma_k(t)).
    \end{split}
\end{equation}
\par Similarly, the data rate of the eavesdropper from MU $k$ can be expressed as

\begin{equation}
\label{eq21}
\begin{split}
    R_{\mathrm{Eve},k}(t) = \log_2(1 + \gamma_{\mathrm{Eve},k}(t)).
\end{split}
\end{equation}
\par Since the eavesdropper can eavesdrop on any of the MU signals according to \cite{DBLP:journals/iotj/QinSWDGYS24, DBLP:journals/tcom/Lv0H023}, the ASR can be written as follows: 

\begin{equation}
\label{eq22}
\begin{split}
    R_k^{\text{sec}}(t) = [R_k(t) - R_{\mathrm{Eve},k}(t)]^+.
\end{split}
\end{equation}
\par The Eq. (\ref{eq22}) ensures that the ASR is always non-negative.
\begin{figure*}[t]
\centering 
\begin{gather}
\label{eq18}
\gamma_k(t) = \frac{\left| \mathbf{w}_k^{1\mathrm{H}} \mathbf{G}(t)^\mathrm{H} \mathbf{h}_{\text{SIM},k}(t) \right|^2 P_k(t)}{
\sum_{j=1, j \neq k}^{K} \left| \mathbf{w}_j^{1\mathrm{H}} \mathbf{G}(t)^\mathrm{H} \mathbf{h}_{\text{SIM},j}(t) \right|^2 P_j(t) + 
\sum_{i=1}^{K} \left| \mathbf{w}_i^{1\mathrm{H}} \mathbf{G}(t)^\mathrm{H} \mathbf{h}_{\text{SIM},i}(t) \right|^2 \kappa_i^2 P_i(t) + N_0}
. \tag{16} \\[10pt] 
\label{eq19}
\gamma_{\mathrm{Eve},k}(t) = \frac{\left| h_{\mathrm{Eve}, k} \right|^2 P_k(t)}{
\sum_{j=1, j \neq k}^{K} \left| h_{\mathrm{Eve}, j} \right|^2 P_j(t) + 
\sum_{i=1}^{K} \left| h_{\mathrm{Eve}, i} \right|^2 \kappa_i^2 P_i(t) + N_0}
.\tag{17}
\end{gather}
\vspace{-6pt}
\noindent\rule{\textwidth}{0.4pt}  
\vspace{-10pt}  
\end{figure*}
\subsection{Problem Formulation}
\label{Problem Formulation for MUs}
\par We formulate a JPPSOP to maximize ASR by jointly optimizing the phase shift $\mathbf{\Phi}_m(t)$ of the SIM and the transmit power $P_k(t)$ of MUs, subject to the minimum ASR requirement $R_{\min}$ for MUs. Specifically, the optimization problem is formulated as follows:

\begin{subequations}
\label{eq:opt_problem}
\begin{align}
    \mathcal{P}: & \max_{\mathbf{\Phi}_m(t), P_k(t)} \quad \frac{1}{T} \sum_{t=1}^T \sum_{k=1}^K [R_k(t) - R_{\mathrm{Eve},k}(t)]^+ \label{eq:opt_problem_a} \\
    & \text{s.t.} \quad {C1: }\; \varphi_m^n(t) \in [0, 2\pi), \forall m \in \mathcal{M}, \forall n \in \mathcal{N}, \label{eq:opt_problem_b} \\
    & \phantom{\text{s.t.}} \quad {C2: }\; 0 \le P_k(t) \le P_{\max}, \forall k\in \mathcal{K}, \forall t \in \mathcal{T},
    \label{eq:opt_problem_c} \\
    & \phantom{\text{s.t.}} \quad {C3: }\; R_k^{\text{sec}}(t) \ge R_{\min}, \forall k\in \mathcal{K}, \forall t \in \mathcal{T}, \label{eq:opt_problem_d}
\end{align}
\end{subequations}

\noindent where ${C1}$ and ${C2}$ impose upper and lower bounds on the decision variables $\mathbf{\Phi}_m(t)$ and $P_k(t)$, respectively, while ${C3}$ ensures that all MUs can achieve a secrecy rate no lower than the threshold within any time slot.
\par In the JPPSOP, both data rate terms in the objective function Eq. (\ref{eq:opt_problem_a}) are logarithmic functions of the SINR, while the SINR itself is a complex fractional function of the optimization variables $\mathbf{\Phi}_m(t)$ and $P_k(t)$. This structure, characterized by the difference of two logarithmic functions and the intricate form of the variables, renders the objective function highly non-concave \cite{DBLP:journals/wcl/NiuAPGCD24, DBLP:journals/ojcs/DarsenaVIG25}. In addition, ${C3}$ is also formulated with complex SINR expressions, resulting in a feasible non-convex region for the JPPSOP. Furthermore, MU introduces dynamic and temporal dimensions to the optimization problem, making it intractable to be solved directly by conventional optimization methods. 

\section{Proposed PPO-BOP}
\label{DRL-Based Approach}
\par In this section, we first elaborate on the motivation for adopting DRL, and then convert JPPSOP into a Markov decision process (MDP). Subsequently, we introduce the PPO and the proposed PPO-BOP with its three enhanced mechanisms. Finally, we analyze the complexity of the proposed PPO-BOP.

\subsection{Motivations for Using DRL} 

\par Due to the properties of the formulated JPPSOP, conventional optimization methods are not suitable for solving this problem. First, the non-linear relationship between the optimization objective and decision variables makes it difficult to find the optimal solution using conventional methods, such as convex optimization \cite{DBLP:journals/twc/YangXZNXW21, DBLP:journals/tcom/PanLSFLY23}. Second, JPPSOP is a long-term and sequential decision-making problem, which means that evolutionary algorithms are not feasible due to uncertainties and dynamic conditions. This necessitates a trade-off between short-term gains and long-term objectives, which in turn leads to the suboptimal performance of the conventional optimization methods \cite{11130648}. Finally, as the number of SIM layers and meta-atoms increased, the solution dimension will be correspondingly increase. The massive solution space suggests that the online algorithms will be inefficient \cite{DBLP:journals/twc/PanLSWY24}.
\par In this context, DRL demonstrates significant advantages for such optimization problems, especially in dynamic environments \cite{DBLP:journals/iotj/QinSWDGYS24, DBLP:journals/twc/ZhangGLXZNN24}, since the core of DRL is the MDP, which provides a mathematical framework for sequential decision-making problems under uncertainty. In an MDP, an agent interacts with the environment at discrete time steps, making decisions to maximize its cumulative reward \cite{10689376}.
Thus, DRL can optimize long-term rewards by considering future outcomes, rather than focusing solely on short-term gains \cite{DBLP:journals/corr/SchulmanWDRK17}. In summary, DRL shows great potential to solve the formulated JPPSOP considering $\mathbf{\Phi}_m(t)$ and $P_k(t)$, handling dynamic environmental changes. 







\subsection{MDP Conversion}
\par In a dynamic mobile environment, JPPSOP can be converted to an MDP. Generally, an MDP is represented as a $5$-tuple $\mathcal{M}_{\text{SIM}} = \langle \mathcal{S}, \mathcal{A}, \mathcal{P}, \mathcal{R}, \Upsilon \rangle$. $\mathcal{S}$ is the state space of the environment. $\mathcal{A}$ is the action space of the agent. $\mathcal{P}$ denotes the probability of state transition of the environment. $\mathcal{R}$ is the reward function. $\Upsilon \in [0,1)$ represents the reward discount factor to balance immediate rewards and future rewards. To solve the problem using DRL-based algorithms, the underlying MDP is defined by the following key elements:
\subsubsection{State Space} 
\par To enable the agent to make effective and secure communication decisions in a complex and dynamic mobile environment, $\mathcal{S}$ must comprehensively capture key information of the environment. In the context of our model, the state information includes the physical channel environment, which is determined by user mobility, and the direct performance feedback from the previous decision. Furthermore, we include the ASR of the individual MU and the overall system as part of the state. This design allows the agent to clearly associate its environment with security performance, learning the optimal policy to configure $\mathbf{\Phi}_m(t)$ and $P_k(t)$ in a dynamic environment. Moreover, including both SINR and ASR vectors enables the agent to simultaneously perceive raw interference patterns and the final security metric, and this explicit feature representation accelerates convergence and stabilizes value estimation. Thus, $\mathcal{S}$ is defined as follows:
\begin{equation}
\label{eq24}
\begin{split}
    \mathbf{s}(t) = [\mathbf{q}(t), \mathbf{r}^{\text{sec}}(t), \bm{\gamma}(t), \bar{r}^{\text{sec}}(t)],
\end{split}
\end{equation}
\noindent where $\mathbf{q}(t)=[q_1(t),...,q_K(t)]$ represents the positions of the MUs.
$\mathbf{r}^{\text{sec}}(t) = [r^{\text{sec}}_1(t), r^{\text{sec}}_2(t), \dots, r^{\text{sec}}_K(t)]$ represents the instantaneous secrecy rate vector of MUs. $\boldsymbol{\gamma}(t) = [\gamma_1(t), \gamma_2(t), \dots, \gamma_K(t)]$ represents the SINR vector of each MU in the BS. $\bar{r}^{\text{sec}}(t) = \frac{1}{K} \sum_{k=1}^{K} r^{\text{sec}}_k(t)$ is system ASR.
\par \subsubsection{Action Space} $\mathcal{A}$ is defined as the set of control parameters available to the agent at the time step. By adjusting the phase shift, the system can intelligently reconfigure the channel conditions to enhance the signals for the MUs while suppressing those for the eavesdropper. In addition, by tuning the power allocation component $P_k$, the system can also effectively manage signal strength and inter-user interference. Therefore, incorporating these two core controllable variables into a unified $\mathcal{A}$ is a natural and necessary choice to consider security performance, which is defined as
\begin{equation}
\label{eq25}
\begin{split}
    \mathbf{a}(t) = [\mathbf{\Phi}(t), \mathbf{P}(t)].
\end{split}
\end{equation}
\par As part of the learning process, the agent learning proceeds step by step. When the agent takes action $\mathbf{a}(t) \in \mathcal{A}$ at the $t$-th time step, the environment changes from the current state $\mathbf{s}(t)$ to the subsequent state $\mathbf{s}(t+1)$.
\par \subsubsection{Reward Function} $\mathcal{R}$ quantifies the immediate reward obtained by taking an action in a given state $\mathbf{s}(t)$, so as to maximize the ASR of the system by executing the action. Specifically, $\mathcal{R}$ is designed as a composite reward function that combines three key components. Its core idea is to reward the actions that improve the overall security performance of the system while satisfying all the constraints. Additionally, the strict constraint checks ensure that valid rewards are only accrued when all users meet the $R_{\min}$. Thus, $\mathcal{R}(s(t),a(t),s(t+1))$ is expressed as follows:
\begin{subequations}
\label{eq:reward}
\begin{gather}
    \mathcal{R} = \mathcal{G}_{\text{diff}} r^{\text{diff}} + \mathcal{G}_{\text{pro}} r^{\text{pro}} - \mathcal{G}_{\text{sta}} r^{\text{sta}}, 
    \label{eq23a} \\
    r^{\text{diff}} \left( \mathbf{s}_t, \mathbf{s}_{t+1} \right) 
    = \bar{r}^{\text{sec}}_{t+1} - \bar{r}^{\text{sec}}_t, 
    \label{eq24a} \\
    r^{\text{pro}} \left( \mathbf{s}_{t+1} \right) 
    = 1 - e^{-\bar{r}^{\text{sec}}_{t+1}}, 
    \label{eq25a} \\
    r^{\text{sta}} \left( \mathbf{s}_t, \mathbf{s}_{t+1} \right) 
    = 
    \begin{cases}
         \left| \bar{r}^{\text{sec}}_{t+1} - \bar{r}^{\text{sec}}_t \right|, & \text{if } \left| \bar{r}^{\text{sec}} \right| > \delta^{\text{sta}} \\
        0, & \text{otherwise}
    \end{cases},
    \label{eq26a}
\end{gather}
\end{subequations}

\noindent where the rate increment reward $r^{\text{diff}}$ encourages the improvement of the ASR. By rewarding this increment, the agent is incentivized to enhance the security performance of the system. The progression reward $r^{\text{pro}}$ is a function of gradual saturating growth, which incentivizes continuous progress and tends to stabilize as ASR increases. Notably, the $r^{\text{pro}}$ at lower rates provides high marginal utility, acting as an implicit soft constraint that naturally drives the system to satisfy $C3$ without requiring abrupt discrete penalties that might destabilize training. Moreover, the stability reward $r^{\text{sta}}$ applies a penalty when the state undergoes drastic changes, reducing the instability of the system and severe oscillations. $\delta^{\text{sta}}$ is the range of rate variation. Crucially, this composite reward design inherently discourages diminishing individual user performance. Additionally, the growth coefficients $\mathcal{G}_{\text{diff}}$, $\mathcal{G}_{\text{pro}}$, and $\mathcal{G}_{\text{sta}}$ are used to amplify the differences, which are determined through extensive empirical validation to balance the trade-off between maximizing ASR and maintaining policy stability. In other words, they act as normalization factors to normalize the varying magnitudes of the reward components, preventing any single term from dominating the gradient estimation.
\begin{figure*}[!t]
\centering
\includegraphics[width=0.99\linewidth]{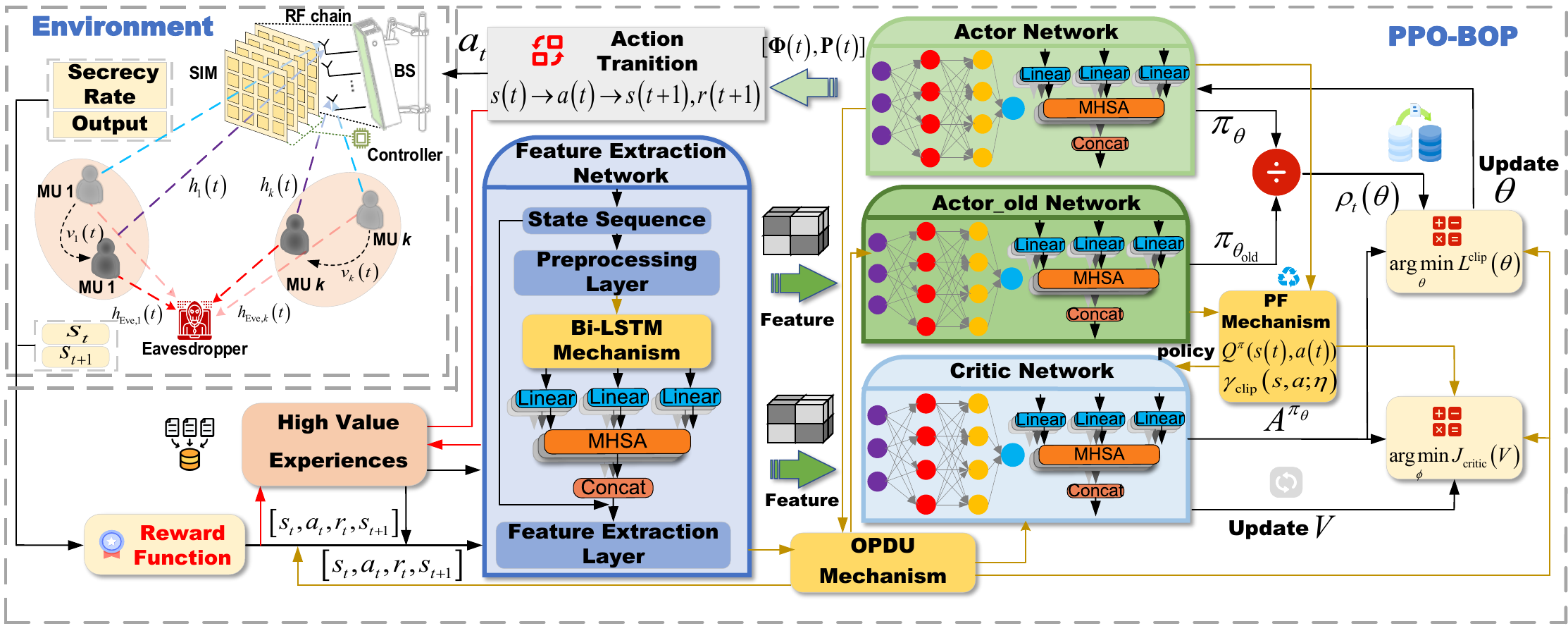}
\caption{The overall architecture of the proposed PPO-BOP to solve the JPPSOP.}
\label{fig2}
\end{figure*}

\subsection{PPO Algorithm}
\par In this work, we adopt PPO as the basic framework. As a policy-based DRL algorithm, PPO generates a policy network $\pi_{\theta}$ to make decisions in the aforementioned MDP \cite{10857476}. Therefore, PPO aims to optimize the policy parameters to achieve higher state values\cite{DBLP:journals/corr/SchulmanWDRK17}, i.e.,
\begin{equation}
\label{eqppo1}
\begin{split}           \textstyle\max_{\theta} J(\theta) = \mathbb{E}_{\tau \sim \pi_{\theta}} \left[ \textstyle\sum_{t=1}^{T} r(t) \right],
\end{split}
\end{equation}
\noindent where $\max_{\theta}$ represents the maximization of the objective function $J(\theta)$ with respect to the policy parameters $\theta$. The trajectory $\tau$ consists of the action of the agent at time step and the feedback. The inefficiency of the sample of the policy gradient methods arises from the need to sample numerous complete trajectories $\tau$, particularly in high-dimensional state and continuous action spaces. To improve efficiency, the actor-critic method was introduced, which uses a critic network to evaluate the effectiveness of actions, and thus can efficiently guide the update of the actor network. To balance the influence of the actor and critic networks during the training process, and to prevent one from dominating the learning, a hyperparameter $\beta_b$ is introduced to balance the loss contributions of the actor and critic, which can be expressed as follows \cite{DBLP:journals/corr/SchulmanWDRK17}:
\begin{equation}
\label{eqppo2}
\begin{alignedat}{1}
    &J(\theta,V) = J_{\text{actor}}(\theta) - \beta_b J_{\text{critic}}(V) \\ 
    &= \mathbb{E}_{\tau \sim \pi_{\theta}}
      \Bigl[\textstyle\sum_{t=1}^{T}\nabla_{\theta}\log \pi_{\theta}(\mathbf{a}(t)\mid \mathbf{s}(t))
      A^{\pi_{\theta}}(\mathbf{s}(t),\mathbf{a}(t))\Bigr] \\ 
    &-\,\mathbb{E}_{\tau}
    \Bigl[\textstyle\sum_{t=1}^{T}\tfrac12\bigl(r(t)+\Upsilon^{t} V(\mathbf{s}(t+1))-V(\mathbf{s}(t))\bigr)^2\Bigr],
\end{alignedat}
\end{equation}
\noindent where $\nabla_{\theta} \log \pi_{\theta}(\mathbf{a}(t) | \mathbf{s}(t))$ is the policy gradient
with respect to the parameter $\theta$. $A^{\pi_{\theta}}(\mathbf{s}(t), \mathbf{a}(t))$ represents the advantage function, which indicates the advantage of taking action $\mathbf{a}(t)$ in state $\mathbf{s}(t)$ compared to the average policy value \cite{10857476}, i.e., 
\begin{equation}
\label{eqppo3}
\begin{split}
    A^{\pi_{\theta}}(\mathbf{s}(t),\mathbf{a}(t)) = Q^{\pi_{\theta}}(\mathbf{s}(t),\mathbf{a}(t)) - V^{\pi_{\theta}}(\mathbf{s}(t)). 
\end{split}
\end{equation}
\par In Eq. (\ref{eqppo2}), $J_{\text{critic}}(V)$ is the loss function of the critic network, which measures the error between the predicted state value and the actual return, to improve the prediction accuracy of the critic network. $V(\mathbf{s}(t))$ is the estimated value of the state $\mathbf{s}(t)$.
$Q^{\pi_{\theta}}(\mathbf{s}(t), \mathbf{a}(t)) = \mathbb{E}^{\pi_{\theta}}[\mathcal{R}(t) | \mathbf{s}=\mathbf{s}(t), \mathbf{a}=\mathbf{a}(t)]$ represents the action value function for state $\mathbf{s}(t)$ and action $\mathbf{a}(t)$.
$V^{\pi_{\theta}}(\mathbf{s}(t)) = \mathbb{E}^{\pi_{\theta}}[\mathcal{R}(t) | \mathbf{s}=\mathbf{s}(t)]$ is the state value function. $A^{\pi_{\theta}}(\mathbf{s}(t),\mathbf{a}(t))$ is obtained by estimating the temporal difference error.
\par Furthermore, PPO uses a clipped surrogate objective function to limit the magnitude of policy updates, thereby stabilizing the training process \cite{DBLP:journals/corr/SchulmanWDRK17}. Its objective function maximizes the advantage function while avoiding too large of policy changes, which can be expressed as follows \cite{DBLP:journals/corr/SchulmanWDRK17}:
\begin{equation}
\label{eqppo4}
\begin{alignedat}{1}
    &L^{\mathrm{clip}}(\theta) ={} \\ 
    &\mathbb{E}_{t}\Bigl[\min\bigl(\rho_t(\theta)A^{\pi_{\theta}}(\mathbf{s}(t),\mathbf{a}(t)),\,
      \rho_t^{\mathrm{clip}}(\theta)A^{\pi_{\theta}}(\mathbf{s}(t),\mathbf{a}(t))\bigr)\Bigr],
\end{alignedat}
\end{equation}
\noindent where $\rho_t(\theta)$ is the probability ratio of the same action in the same state and $\rho_t^{\text{clip}}(\theta) = \text{clip}(\rho_t(\theta), 1-\epsilon, 1+\epsilon)$ is the clipped value of $\rho_t(\theta)$, with $\epsilon$ being a hyperparameter that controls the clipping range.
\par Inspired by \cite{DBLP:journals/iotj/AungNTHH24}, the multi-head self-attention (MHSA) mechanisms is embedded to enhance the performance of PPO. A standard fully connected network may struggle to effectively capture these critical internal dependencies when processing such inputs. To overcome this, the MHSA mechanism is embedded into both the actor and critic networks of the PPO, which enables the model to automatically focus on the environmental information most critical to the current decision, enhances the modeling of dynamic correlations such as user mobility and channel variations. Furthermore, the MHSA effectively disentangles complex inter-MU interference patterns by attending to different state spaces simultaneously. In summary, MHSA complements the temporal processing capabilities of PPO.
\par However, within the converted MDP, the environmental feedback is characterized by substantial delays and severe fluctuations, making it difficult for PPO to capture effective long-term dependencies in a short period. Furthermore, PPO entirely relies on online interaction data, leading to low sample efficiency. In addition, PPO often incurs significant policy evaluation bias during policy optimization, struggling to strike a balance between stability and generalization capabilities. Thus, we embed three enhanced mechanisms into PPO so that it can better align with our converted MDP.

\subsection{PPO-BOP}
\par To address the challenges faced by PPO in solving our converted MDP, we propose PPO-BOP with three enhanced mechanisms, and their advantages are as follows:
\begin{itemize}
\item \textit{Bi-LSTM Mechanism:} Bi-LSTM mechanism enables forward-looking decision-making using historical channel state information and user mobility trends. Consequently, it achieves a more stable optimization of $\mathbf{\Phi}_m(t)$ and $P_k(t)$ in dynamic environments.
\item \textit{OPDU Mechanism:} OPDU mechanism enables the agent to quickly learn an effective policy for joint power control and phase shift configuration, allowing it to adapt to time-varying channel conditions without requiring extensive real-time interaction with the environment.
\item \textit{PF Mechanism:} PF mechanism allows the critic network to promptly track the actor network updates of $\mathbf{\Phi}_m(t)$ and $P_k(t)$ policies in response to user mobility, thereby enabling proactive self-adaptation to dynamic channel conditions. This ensures that the PPO-BOP can consistently and reliably optimize the ASR, even in dynamic environments.
\end{itemize}
\par As shown in Fig. \ref{fig2}, the PPO-BOP architecture consists of three parts: environment interaction, feature extraction, and a decision network. The agent uses a feature extraction network with a Bi-LSTM to perceive and process the environmental state. Then, the core actor-critic decision network incorporates an innovative fusion of OPDU and PF mechanism to generate optimal $\mathbf{\Phi}_m(t)$ and $P_k(t)$ policies with higher efficiency. To tackle the high dimensional continuous action space inherent in SIM control, this architecture leverages PPO linear scaling Gaussian policy, where Bi-LSTM simplifies policy mapping via feature compression, OPDU enhances exploration efficiency through data reuse, and PF stabilizes value estimation against high variance updates. The pseudo-code of PPO-BOP is given in Algorithm \ref{alg:ppo-bop}, and the details of the enhanced mechanisms are presented below.
\subsubsection{Bi-LSTM Mechanism} 
\par As previously discussed, user mobility and time-varying channel fading introduce significant dynamics to the considered system and the MDP, which can reduce learning accuracy and increase complexity. Furthermore, the current direction and velocity of an MU typically depend on its previous ones, while the conventional PPO structure has difficulty capturing such temporal dependencies in sequential observations \cite{10857476}. While a single state vector in the defined state space essentially functions as a static snapshot, the temporal sequence of historical states can be utilized to reveal implicitly encoded dynamic information. Therefore, we incorporate a Bi-LSTM into the PPO-BOP feature network to enhance its temporal processing and reasoning capabilities. Specifically, Bi-LSTM uses historical channel information and user mobility trends to make forward-looking decisions, thereby achieving a more stable optimization of $\mathbf{\Phi}_m(t)$ and $P_k(t)$ in dynamic environments \cite{DBLP:journals/iotj/HeZWLGG23}. Structurally, the Bi-LSTM is positioned within the shared feature extraction layer preceding both the actor and critic networks. This placement ensures that the encoded temporal dependencies are provided as a common context for both action selection and value estimation.
\par Moreover, an LSTM unit consists of an input gate, an output gate, and a forget gate, which enable the algorithm to manage temporal dependencies by selectively filtering information at time step. Mathematically, the operation of an LSTM unit can be described as follows \cite{DBLP:journals/iotj/HeZWLGG23}:
\begin{subequations}
\label{eq:lstm}
\begin{align}
    f(t) &= \sigma(W_f \cdot [l(t-1), x(t)] + b_f), \label{eqlstm2} \\
    i(t) &= \sigma(W_i \cdot [l(t-1), x(t)] + b_i), \label{eqlstm3} \\
    \hat{C}(t) &= \tanh(W_C \cdot [l(t-1), x(t)] + b_C), \label{eqlstm4} \\
    C(t) &= f(t) * C(t-1) + i(t) * \hat{C}(t), \label{eqlstm5} \\
    o(t) &= \sigma(W_o \cdot [l(t-1), x(t)] + b_o), \label{eqlstm6} \\
    l(t) &= o(t) * \tanh(C(t)), \label{eqlstm7}
\end{align}
\end{subequations}
\noindent where $x(t)$ is the input vector at the current time step $t$. $l(t-1)$ is the output (or hidden state) of the previous time step $(t-1)$. $f(t)$, $i(t)$, and $o(t)$ are the activations of the forget, input, and output gates, respectively. $W_f$, $W_i$, $W_C$, $W_o$ are the weight matrices for each respective layer. $b_f$, $b_i$, $b_C$, $b_o$ are the bias vectors for each respective layer. $\hat{C}(t)$ is the candidate cell state. $C(t)$ is the cell state at timestep $t$. $\sigma$ and $\tanh$ are the activation functions of the sigmoid and hyperbolic tangent, respectively. Bi-LSTM mechanism uses two LSTMs to learn forward and backward information, respectively, capturing a representation of bidirectional information \cite{DBLP:journals/iotj/HeZWLGG23}. Therefore, the forward hidden layer $\overrightarrow{l(t)}$ processes the sequence starting from the first token, while the backward hidden layer $\overleftarrow{l(t)}$ processes it starting from the last token. The forward and backward internal states of the Bi-LSTM mechanism can be summarized as follows \cite{DBLP:journals/iotj/AungNTHH24}:
\begin{equation}
\label{eqlstm8}
\begin{aligned}
    \overrightarrow{l(t)} &= \overrightarrow{\text{LSTM}}(x(t), l(t-1)), \\
    \overleftarrow{l(t)} &= \overleftarrow{\text{LSTM}}(x(t), l(t-1)).
\end{aligned}
\end{equation}
\par Then, the final output is jointly determined by the hidden layers from both directions, and is represented as \cite{DBLP:journals/iotj/AungNTHH24}
\begin{equation}
\label{eqlstm9}
\begin{split}
    y(t) = \overrightarrow{l(t)} + \overleftarrow{l(t)}.
\end{split}
\end{equation}
\par This bidirectional feature extraction compensates for the lack of explicit velocity parameters in the raw state space, enabling accurate prediction of next state transitions. By capturing temporal dependencies in user mobility and channel conditions, the Bi-LSTM mechanism helps PPO-BOP learn effective policies more quickly in dynamically changing environments \cite{10857476}. Moreover, we further enhance Bi-LSTM mechanism by integrating a residual structure to improve the extraction capacity of state information \cite{DBLP:journals/iotj/AungNTHH24}. The Bi-LSTM mechanism is uniquely adapted to transform historical channel observations into predictive adjustments of the phase shift $\mathbf{\Phi}_m(t)$, directly addressing the decision latency in highly dynamic environments.
\par \subsubsection{OPDU Mechanism} In considering the time-varying channel conditions characterized by MUs, PPO uses online interaction data for policy updates, which leads to low sample efficiency and limited policy adaptation speed under such complex and dynamic conditions. For this purpose, we employ the OPDU mechanism that transforms PPO-BOP from an online learning algorithm into a hybrid learning framework by introducing the capability of utilizing offline data. This approach addresses the challenges faced by PPO in highly dynamic mobile communication environments, which ultimately allows efficient and rapid learning of the control strategies for $\mathbf{\Phi}_m(t)$ and $P_k(t)$. Specifically, the offline dataset is dynamically populated into a replay buffer during the concurrent online training, implying that the behavior policy $\mu$ represents a mixture of the agent evolving historical policies. To ensure data quality, we employ a prioritization strategy that selectively retains high value experiences yielding rewards exceeding a dynamic threshold \cite{DBLP:journals/tits/ChenATCWBG25}, where the high value experiences are defined as transition tuples yielding rewards that exceed a dynamic moving average baseline, representing successful configurations of $\mathbf{\Phi}_m(t)$ and $P_k(t)$ with exceptional ASR. By maintaining this dynamic threshold, the system ensures that the most influential historical data is preserved. We define the surrogate objective for offline data \cite{meng2023off}:
\begin{equation}
\label{eqOPDU1}
\begin{split}
    L_{\mu}(\pi) = \mathbb{E}_{s \sim \rho\mu, a \sim \mu} \left[ \frac{\pi(\mathbf{a}(t)|\mathbf{s}(t))}{\mu(\mathbf{a}(t)|\mathbf{s}(t))} A_{\pi_{\text{old}}}(\mathbf{s}(t),\mathbf{a}(t)) \right],
\end{split}
\end{equation}
\noindent where $\pi_{\text{old}}$ is the current policy. $\rho^{\mu}$ represents the distribution of state visits under the behavior policy $\mu$. The term $A_{\pi_{\text{old}}}(\mathbf{s}(t),\mathbf{a}(t))$ denotes the advantage of taking action $\mathbf{a}$ in state $\mathbf{s}$, relative to the average action, under the current policy $\pi_{\text{old}}$. Moreover, to adapt to the discrepancy between the behavior policy $\mu$ and the target policy, OPDU modifies the clipping limits to effectively use offline collected data under $\mu$. Concretely, Its objective function is defined as:
\begin{equation}
\label{eqOPDU2}
\begin{alignedat}{1}
    &L_{\mathrm{Off\text{-}Policy\ PPO}}^{\text{clip}}(\pi) = \\
    &\mathbb{E}_{s\sim\rho^\mu,\,a\sim\mu}
      \Bigl[\min\bigl(
        r_{\pi}(\mathbf{s}(t),\mathbf{a}(t))\,A_{\pi_{\mathrm{old}}}(\mathbf{s}(t),\mathbf{a}(t)),\\
    &\quad
        \mathrm{clip}\bigl(r_{\pi}(\mathbf{s}(t),\mathbf{a}(t)),\,l_{\mathbf{s}(t),\mathbf{a}(t)},\,h_{s(t),a(t)}\bigr)
        \,A_{\pi_{\mathrm{old}}}(\mathbf{s}(t),\mathbf{a}(t))
      \bigr)\Bigr],
\end{alignedat}
\end{equation}
\noindent where $r_{\pi}(\mathbf{s}(t),\mathbf{a}(t)) = \frac{\pi(\mathbf{a}(t)|\mathbf{s}(t))}{\mu(\mathbf{a}(t)|\mathbf{s}(t))}$ is the policy ratio and $\mu(\mathbf{a}(t)|\mathbf{s}(t))$ is the behavior policy. $l_{\mathbf{s}(t),\mathbf{a}(t)} = \frac{\pi_{old}(\mathbf{a}(t)|\mathbf{s}(t))}{\mu(\mathbf{a}(t)|\mathbf{s}(t))}(1-\epsilon)$ is the adjusted clipped lower bound.
$h_{\mathbf{s}(t),\mathbf{a}(t)} = \frac{\pi_{old}(\mathbf{a}(t)|\mathbf{s}(t))}{\mu(\mathbf{a}(t)|\mathbf{s}(t))}(1+\epsilon)$ is the upper bound adjusted with the clipped.
\par To control the bias introduced by offline data, we introduce a kullback–leibler (KL) divergence-based adaptive adjustment mechanism. Specifically, the KL divergence between the current policy $\pi$ and the behavior policy $\mu$  can be calculated by \cite{meng2023off}
\begin{equation}
\label{eqOPDU3}
\begin{split}
    D_{\text{KL}}(\mu || \pi) = \mathbb{E}_{s \sim \rho_{\mu}, a \sim \mu} \left[ \log \frac{\mu(\mathbf{a}(t)|\mathbf{s}(t))}{\pi(\mathbf{a}(t)|\mathbf{s}(t))} \right].
\end{split}
\end{equation}
\par Due to the different magnitude of the KL divergence, we dynamically adjust the proportion of offline data usage by introducing an adaptive coefficient $\alpha_{\text{KL}}$, which is given by \cite{meng2023off}
\begin{equation}
\label{eqOPDU4}
\begin{split}
    \alpha_{\text{KL}} = \begin{cases} \max(\alpha_{\min}, \alpha_{\text{KL}} \cdot \frac{\delta_{\text{KL}}}{D_{\text{KL}}}), & \text{if } D_{\text{KL}} > \delta_{\text{KL}}, \\ \min(1.0, \alpha_{\text{KL}} \cdot 1.05), & \text{otherwise}, \end{cases}
\end{split}
\end{equation}
\noindent where $\delta_{\text{KL}}$ is the KL divergence threshold and $\alpha_{\min}$ is the minimum coefficient value. According to Eq. (\ref{eqOPDU4}), the contribution of offline data diminishes with large KL divergence, while it increases when the divergence remains within an acceptable threshold. The OPDU mechanism represents a specific modification for non-stationary wireless environments, where the dynamic KL-divergence adjustment prevents the distribution shifts caused by MU mobility. Moreover, it mainly relies on off-policy correction methods, such as value function trace estimation and advantage calculation. These techniques aim to reduce the gap between the historical behavior policy that produced the data and the target policy being optimized, and provide a stable and low-variance method to accurately estimate the value and advantage of the target policy. Consequently, this accurate estimation facilitates more consistent and reliable policy updates toward the optimal power and phase shift strategy, significantly enhancing sample efficiency, and ensuring stability \cite{meng2023off}.
\par \subsubsection{PF Mechanism} Similarly, dynamic channel conditions create significant temporal mismatch and bias between policy updates and value estimation. However, the critic network in conventional PPO relies solely on environmental rewards for its updates, failing to promptly reflect the impact of policy changes on future security performance. This leads to slow policy convergence and low learning efficiency, and we introduce a PF mechanism that enables the critic network to instantly capture the adjustments made by the actor network, i.e., efficiently adjusting $\mathbf{\Phi}_m(t)$ and $P_k(t)$ according to the user locations, thereby achieving proactive self-adaptation to the dynamic channel. Specifically, the update of the critic network not only depends on the rewards provided by the environment, but also incorporates feedback based on the policy probabilities output by the actor network.
\par Assuming that the probability of the output of the actor network $\pi(\mathbf{a}(t)|\mathbf{s}(t))$ and the estimate of the value function of the critic network is $V^\pi(\mathbf{s}(t))$, update of the critic network with the proposed PF is given by \cite{DBLP:journals/tsmc/GuCCW22}
\begin{equation}
\label{eqPF1}
\begin{split}
    V^\pi(\mathbf{s}(t)) = \mathbb{E}_\pi[\hat{R}_{\text{PBE}}(t)|\mathbf{s}(t)],
\end{split}
\end{equation}
\noindent where $\hat{R}_{\text{PBE}}(t)$ is the policy-based expected (PBE) value, representing the sum of future discounted rewards with each reward being weighted by the probability of the corresponding action sequence under policy $\pi$. It is defined as follows
\cite{DBLP:journals/tsmc/GuCCW22}:
\begin{equation}
\label{eqPF2}
\begin{split}
    \hat{R}_{\text{PBE}}(t) = \textstyle\sum_{c=t}^{T} \Upsilon^{c-t} r(c) \textstyle\prod_{v=t}^{c} \pi(\mathbf{s}(v), \mathbf{a}(v)),
\end{split}
\end{equation}
\noindent where $\Upsilon$ is the discount factor, $r(c)$ is the reward, and $\pi(\mathbf{s}(v), \mathbf{a}(v))$ is the probability of policy selection action $\mathbf{a}(v)$ in state $\mathbf{s}(v)$. To improve the sensitivity of the value function to policy changes, the PBE value function directly incorporates the influence of the policy on future actions, thus avoiding the problem of decoupling between the policy and the value estimation. When taking action $\mathbf{a}(t)$ in state $\mathbf{s}(t)$, the PBE value function $Q^{\pi}(\mathbf{s}(t), \mathbf{a}(t))$ can be represented as follows \cite{DBLP:journals/tsmc/GuCCW22}:
\begin{multline}
\label{eqPF3}
    Q^{\pi}(\mathbf{s}(t),\mathbf{a}(t)) = r(t) + \\
    \shoveleft
    \Upsilon \sum_{\mathbf{a}(t+1)} 
      \pi\bigl(\mathbf{s}(t+1),\mathbf{a}(t+1)\bigr) Q^{\pi}\bigl(\mathbf{s}(t+1),\mathbf{a}(t+1)\bigr),
\end{multline}
\noindent where $r(t)$ is the immediate reward, $\pi(\mathbf{s}(t+1), \mathbf{a}(t+1))$ is the probability of selecting action $\mathbf{a}(t+1)$ in state $\mathbf{s}(t+1)$ according to policy, and $Q^{\pi}(\mathbf{s}(t+1), \mathbf{a}(t+1))$ is the value of the future pair of state action. This design enables the value function to better reflect changes in the policy, thereby improving the accuracy of the value estimation. To accurately assess the benefits of an action with this PF mechanism \cite{DBLP:journals/tsmc/GuCCW22}, we compute our final advantage function using generalized advantage estimation with the PBE-based value estimates.
The PF mechanism efficiently addresses the complex coupling in multi-layer SIMs, enabling the critic to evaluate phase configurations even when immediate ASR rewards are obscured by channel noise.
\par Ultimately, these mechanisms simultaneously address user mobility and the non-convexity of the problem. Specifically, the Bi-LSTM extracts predictive temporal features that enable the PF mechanism to stabilize value estimation against channel fluctuations. This stability, in turn, allows the OPDU to aggressively reuse historical data for rapid convergence while mitigating off-policy instability.

\begin{algorithm}[tbp]
\caption{PPO-BOP}
\label{alg:ppo-bop}
\KwIn{Actor network $\pi_\theta$, Critic network $V_\phi$, Feature network $F_\psi$}
\KwOut{Optimized $\mathbf{\Phi}_m(t)$ and $P_k(t)$}

Initialize networks $\pi_\theta$, $V_\phi$, $F_\psi$, offline data from $\mu$, high value experiences\;

\For{each episode}{
    Reset environment, get initial state $s_0$, initialize LSTM states $l_o, C_o$\;
    
    \For{time step $t = 1$ \KwTo $T$}{
        \tcp{\textbf{Bi-LSTM Feature Extraction}}
        Process bidirectional states $\overrightarrow{l(t)}, \overleftarrow{l(t)}$ and extract $f_t = y(t)$ via {Eq. (\ref{eqlstm8})-(\ref{eqlstm9})};
        
        Sample action $a_t = [\mathbf{\Phi}_m(t), P_k(t)] \sim \pi_\theta(f_t)$\;
        Execute $a_t$ by observe $s_{t+1}, r_t$ via Eq. (\ref{eq23a})\;
        
Store $(s_t, a_t, r_t, s_{t+1})$ in replay buffer for offline data $\mu$ and prioritizing high value experiences}\;

    \If{$|\mathcal{\mu}| \geq$ batch size}{
        \tcp{\textbf{OPDU Data Utilization}}
        Sample mixed batch from offline data $\mu$ and current policy $\pi$ data\;
        Compute surrogate objective $L_\mu(\pi)$ for offline data via Eq. (\ref{eqOPDU1})\;
        Compute $D_{KL}(\mu||\pi)$ and adjust $\alpha_{KL}$ via Eq. (\ref{eqOPDU3})-(\ref{eqOPDU4})\;
        
        \For{each update epoch}{
            \tcp{\textbf{PF Value Estimation}}
            Compute $\hat{R}_{\text{PBE}}(t)$ via Eq. (\ref{eqPF2})\;
            Update critic network $V^\pi(s(t))$ via Eq. (\ref{eqPF1})\;
            
            Compute advantages with GAE using PBE-based value estimates\;
            Compute actor loss $L^{\text{clip}}(\theta)$ via Eq. (\ref{eqppo4})\;
            
            Update networks $(\theta, \phi, \psi)$ via gradient descent with respective learning rates\;
        }
    }
}

\Return{Optimized policy $\pi_\theta^*$ for $\mathbf{\Phi}_m(t)$ and $P_k(t)$.}
\end{algorithm}

\subsection{Complexity Analysis}
\par This section analyzes the computational and spatial complexities of PPO-BOP during the training phase and the execution phase.
\par \subsubsection{Training Phase} Assume that the total number of training steps is $\mathcal{Z}_{\text{epochs}}$, and a network update is performed every $U$ steps. Then, the total computational complexity (CC) of the training phase is $\mathcal{O}\left(\mathcal{Z}_{\text{epochs}}(\Gamma) + \frac{\mathcal{Z}_{\text{epochs}}}{U} (B \log B + bE(\Gamma + hf_c + f_c))\right)$. It can be summarized as follows:
\begin{itemize}
\item {\textit{Network Initialization}}: Assume that the dimension of the hidden layer of Bi-LSTM is $h$, the number of layers is $l$, and the number of heads in the attention mechanism is $a$. The CC of initialization is $\mathcal{O}(n_s h + h^2 l + ha)$, where $n_s$ is the state dimension. The CC initialization of the actor network and the critic network is $\mathcal{O}(hf_a + f_a n_a)$ and $\mathcal{O}(hf_c + f_c)$, respectively, where $f_a$ and $f_c$ are the hidden feature dimensions of the actor and critic, respectively, and $n_a$ is the dimension of the action space.
\item {\textit{Action Sampling}}: The CC of feature extraction is $\mathcal{O}(n_s h + h^2 l + ha)$. The CC of forward propagation of the actor network is $\mathcal{O}(hf_a + f_a n_a)$. The CC of action sampling is $\mathcal{O}(n_a)$. The total CC of action sampling is $\Gamma = \mathcal{O}(n_s h + h^2 l + ha + hf_a + f_a n_a + n_a)$.
\item {\textit{Replay Buffer Collection}}: The CC of the PPO-BOP add operation is $\mathcal{O}(1)$, while the CC of the update priority is $\mathcal{O}(\log B)$, where $B$ is the buffer size. The CC of the replay buffer collection is $\mathcal{O}(1+\log B)$.
\item {\textit{Network Update}}: The CC of prioritized sampling is $\mathcal{O}(B \log B)$. The CC of the target of the V-trace and the estimation of the advantages is $\mathcal{O}(b)$, where $b$ is the batch size. The CC of the KL divergence and the adjustment priority is $\mathcal{O}(b)$. The update CC of the feature network, the actor network and the critic network is $\mathcal{O}(b(\Gamma + hf_c + f_c))$. The number of iterations is $E$. Therefore, the total CC of each network update is $\mathcal{O}(B \log B + bE(\Gamma + hf_c + f_c))$.
\end{itemize}
\par The space complexity of the training phase space includes mainly the storage of model parameters and the replay buffer of the experience. The total space complexity is $\mathcal{O}(\Gamma + hf_c + f_c + B(n_s + n_a + 1))$.
\par \subsubsection{Execution Phase} Only network forward propagation is needed, without parameter updates and experience storage. The CC of the execution phase mainly includes feature extraction and action generation, which is $\mathcal{O}(\Gamma)$. The execution phase only needs to store model parameters, not to maintain the replay buffer and optimizer state. The space complexity of the execution phase is $\mathcal{O}(\Gamma + hf_c + f_c)$.

\section{Simulation Results and Analysis}
\label{Simulation results and analysis}

\par In this section, we present extensive simulation results to evaluate the performance of PPO-BOP. Without loss of generality, we consider the service area to be a 100\,m $\times$ 100\,m square region. The BS and the eavesdropper are located at coordinates (0\,m, 0\,m, 20\,m) and (20\,m, 20\,m, 0\,m), respectively. MUs are randomly distributed within the area. We assume the thickness of the SIM is $T_{\text{SIM}} = 5\lambda$, ensuring that the spacing between two adjacent metasurfaces in the $M$-layer SIM is $d_{\text{Layer}} = T_{\text{SIM}}/M$ \cite{shi2025downlink}. In the proposed PPO-BOP, the parameters of the actor network and the critic network are updated using the AdamW optimizer. The clipping parameter is set as 0.3. Furthermore, the number of task iterations for the warm-up phase and the evolutionary phase is configured to 50 and 500, respectively. Other specific system parameters are summarized in Table \ref{tab:sim_params}.
\begin{table}[tbp]
\centering
\caption{Simulation Parameters}
\label{tab:sim_params}
\begin{tabularx}{\linewidth}{c l X} 
\toprule
& Parameter & Value \\
\midrule
\multirow{8}{*}{\rotatebox[origin=c]{90}{Scenario}} 
& Number of MUs $k$ & 2 \\
& Maximum velocity $V_{\max}$ & 2 m/s \cite{DBLP:journals/tifs/GuoJCW24}\\
& Path loss exponent & 2.0 \cite{10857476}\\
& Carrier frequency  & 3.5 GHz \cite{10556753} \\
& AWGN power $N_0$ & -110 dBm/Hz \cite{shi2025downlink}\\
& Rician factor $\xi$ & 10 dB \cite{DBLP:journals/tifs/GuoJCW24}\\
& Path loss at 1 m $\beta$ & -20 dB \cite{shi2025downlink}\\
\midrule
\multirow{7}{*}{\rotatebox[origin=c]{90}{PPO-BOP}}
& Convergence threshold $\mathcal{E}$ & $10^{-2}$ \cite{DBLP:journals/twc/YangXZNXW21}\\
& Number of training epochs & 2000 \cite{DBLP:journals/tmc/ZhangSLWWNL25}\\
& Number of time slots per epoch & 40 \cite{DBLP:journals/tmc/ZhangSLWWNL25}\\
& Duration of each time slot $\Delta t$ & 1 s \cite{DBLP:journals/tifs/GuoJCW24}\\
& Discount factor $\Upsilon$ & 0.98 \cite{meng2023off}\\
& Target KL divergence & 0.02 \cite{meng2023off}\\
& Clipped discount factor & 0.7 \cite{DBLP:journals/tsmc/GuCCW22}\\
& Experience replay buffer & $10^6$ \cite{DBLP:journals/tits/ChenATCWBG25}\\
& KL divergence threshold $\delta_{KL}$ & 0.5 \cite{DBLP:journals/tits/ChenATCWBG25}\\
& Probability weighting coefficient & 0.7 \cite{DBLP:journals/tsmc/GuCCW22}\\
\bottomrule
\end{tabularx}
\end{table}
\subsection{Comparison with Other Strategy} 
To evaluate the effectiveness of PPO-BOP, we compared it with the following benchmarks:
\begin{itemize}
\item {\textit{(Strategy 1) Single-antenna System without SIM}}: In this strategy, the BS operates without an SIM, using a traditional single antenna to directly receive signals from the MUs, and the MU transmit power is set as the maximum level.


\item {\textit{(Strategy 2) Uniform SIM Phase Shift and Transmit Power}}: In this strategy, the SIM phase shift of all metasurface elements is set as the fixed value $\pi$. Meanwhile, the transmit power for each MU is set as $P_{\max}/2$.

\item {\textit{(Strategy 3) Uniform SIM Phase and Maximum Transmit Power}}: In this strategy, the SIM phase is set as $\pi$, while each MU transmit power is set as the maximum level.
\end{itemize}
\par
\begin{figure}[htbp]
\centering
\includegraphics[width=0.40\textwidth]{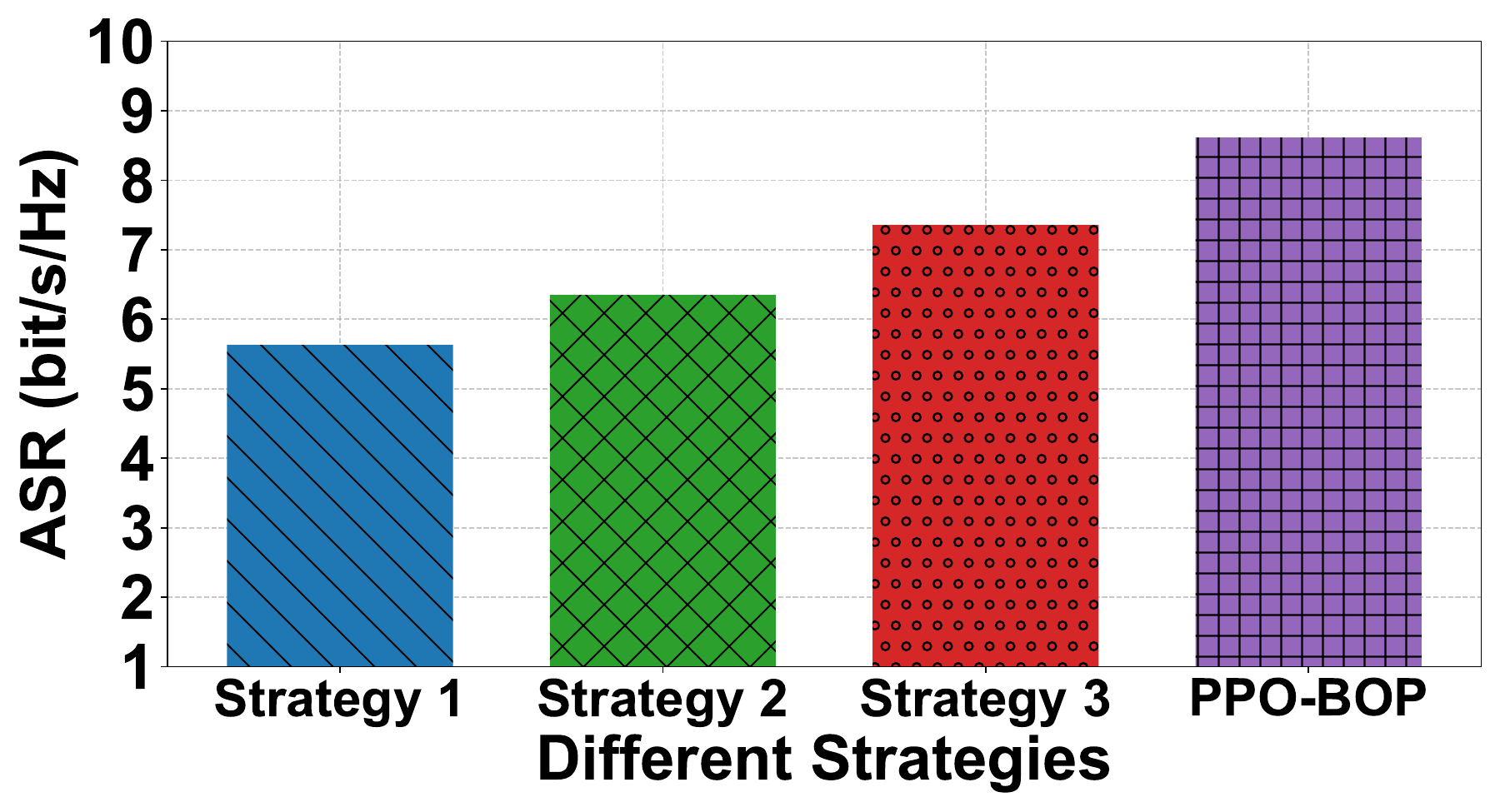}
\caption{ASR when adopting different strategies ($M=4$, $N=36$, $P_{\max}=30$ dBm, $R^{\min}=0.5$ bit/s/Hz, $\kappa=0.10$, $lr=0.001$, $b=128$, and $l=3$, where $lr$, $b$, and $l$ represent the learning rate, the batch size, and the number of Bi-LSTM layers of PPO-BOP, respectively).}
\label{fig3}
\end{figure}

 Fig. \ref{fig3} shows that PPO-BOP outperforms other strategies. Specifically, the comparison with Strategy 1 reveals the necessity of introducing SIM, since SIM provides additional spatial degrees of freedom to enhance ASR. Moreover, the comparison with Strategies 2 suggests the importance of real-time phase shift and transmit power control, since static strategies are unable to adapt to the channel time-variability induced by user mobility. In addition, the comparison with Strategy 3 indicates that, although the MUs under our proposed PPO-BOP do not always transmit at maximum power, the flexible and real-time adaptation of the phase shift still enables the system to achieve a higher ASR.
\subsection{Comparison with Baseline DRL Algorithms} 
\par To evaluate the effectiveness of PPO-BOP, we compare its performance against several DRL baselines, including the double deep Q-network (DDQN) \cite{DBLP:journals/twc/YangXZNXW21}, the deep deterministic policy gradient (DDPG) \cite{DBLP:journals/iotj/QinSWDGYS24}, the twin delayed deep deterministic policy gradient (TD3) \cite{DBLP:journals/twc/ZhangGLXZNN24}, soft actor-critic (SAC) \cite{10689376}, and PPO \cite{DBLP:journals/corr/SchulmanWDRK17}. In fairness, all parameters of these algorithms are kept consistent with those of PPO-BOP.
\par
\begin{figure}[htbp]
\centering
\includegraphics[width=0.49\textwidth]{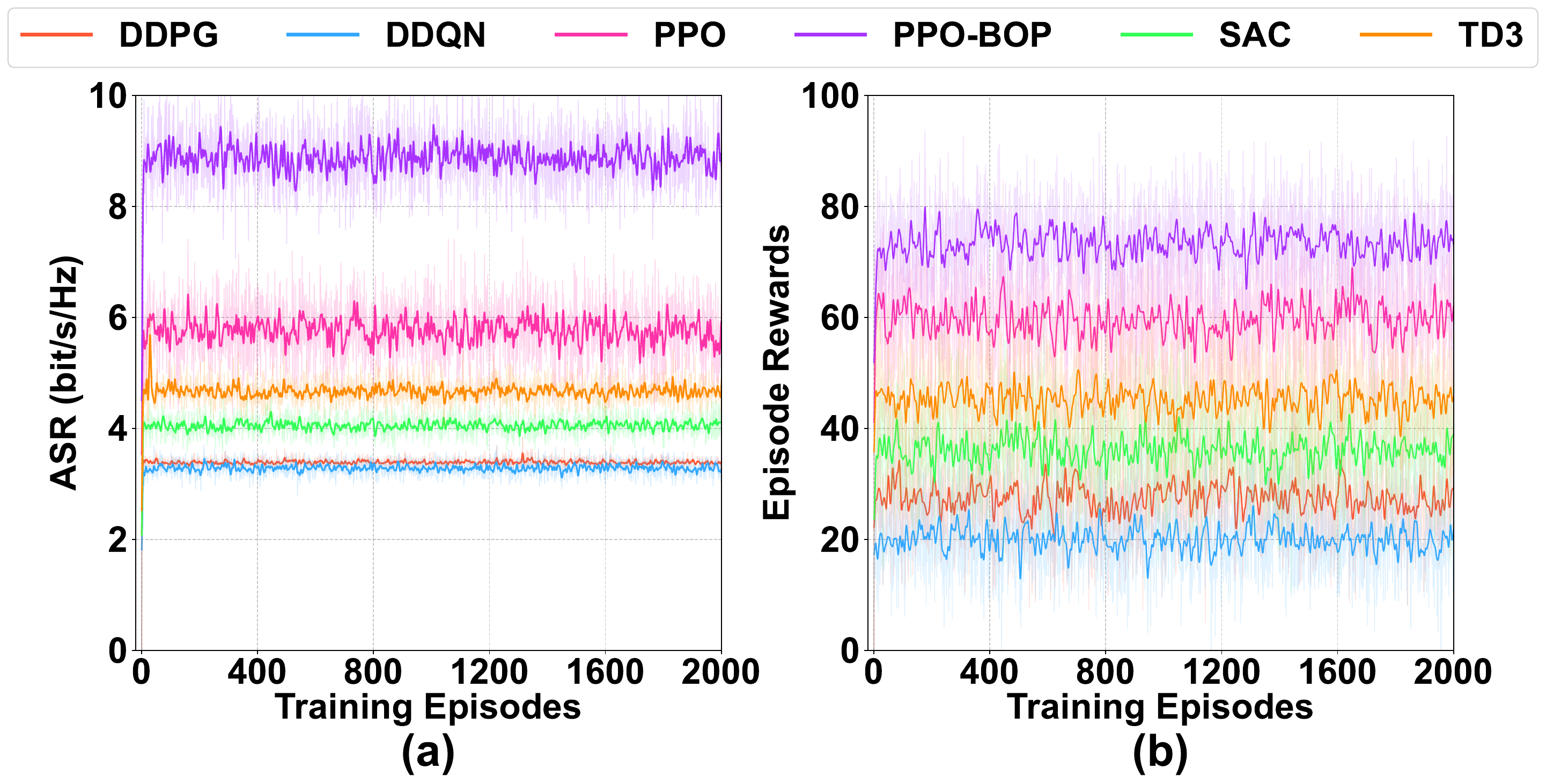}
\caption{Comparison of convergence curves between the proposed PPO-BOP and baseline algorithms. (a) ASR over training episodes. (b) Average reward per episode during training. ($M=4$, $N=36$, $P_{\max}=30$ dBm, $R^{\min}=0.5$ bit/s/Hz, $\kappa=0.10$, $lr=0.001$, $b=128$, and $l=3$).}
\label{fig_sim}
\end{figure}

 Fig. \ref{fig_sim}(a) depicts the evolution of the ASR for PPO-BOP and the baseline DRL algorithms over the training episodes. All algorithms show a clear upward trend during the initial stage, indicating that the algorithms gradually optimize the phase shift and the transmit power through the learning process. As can be seen, after an initial learning phase, the ASR of PPO-BOP rapidly increases and stabilizes in the high range of 8.5-9.0 bit/s/Hz, significantly exceeding the baseline DRL algorithms, which can be explained by the fact that by leveraging Bi-LSTM mechanism to capture temporal features of user mobility and channel variations. Additionally, by integrating the OPDU and PF mechanism for rapid convergence and adaptation, PPO-BOP achieves a higher and more stable ASR in dynamic environments.
\par Fig. \ref{fig_sim}(b) illustrates the reward per episode during the training process for the proposed PPO-BOP compared to other DRL algorithms. All algorithms exhibit a rapid increase in reward values during the early stages of training, after which the reward curves gradually converge, indicating that the learned policies have reached a relatively stable level. PPO-BOP demonstrates superior performance, as its reward curve converges to the highest level among all algorithms, stabilizing between approximately 78 and 80, with relatively small fluctuations after convergence. In summary, PPO-BOP achieves higher rewards, faster convergence, and greater stability, a performance advantage stemming from its use of a PF and Bi-LSTM mechanism, which enhances temporal information processing and thereby accelerates convergence.
\par
\begin{figure}[htbp]
\centering
\includegraphics[width=0.42\textwidth]{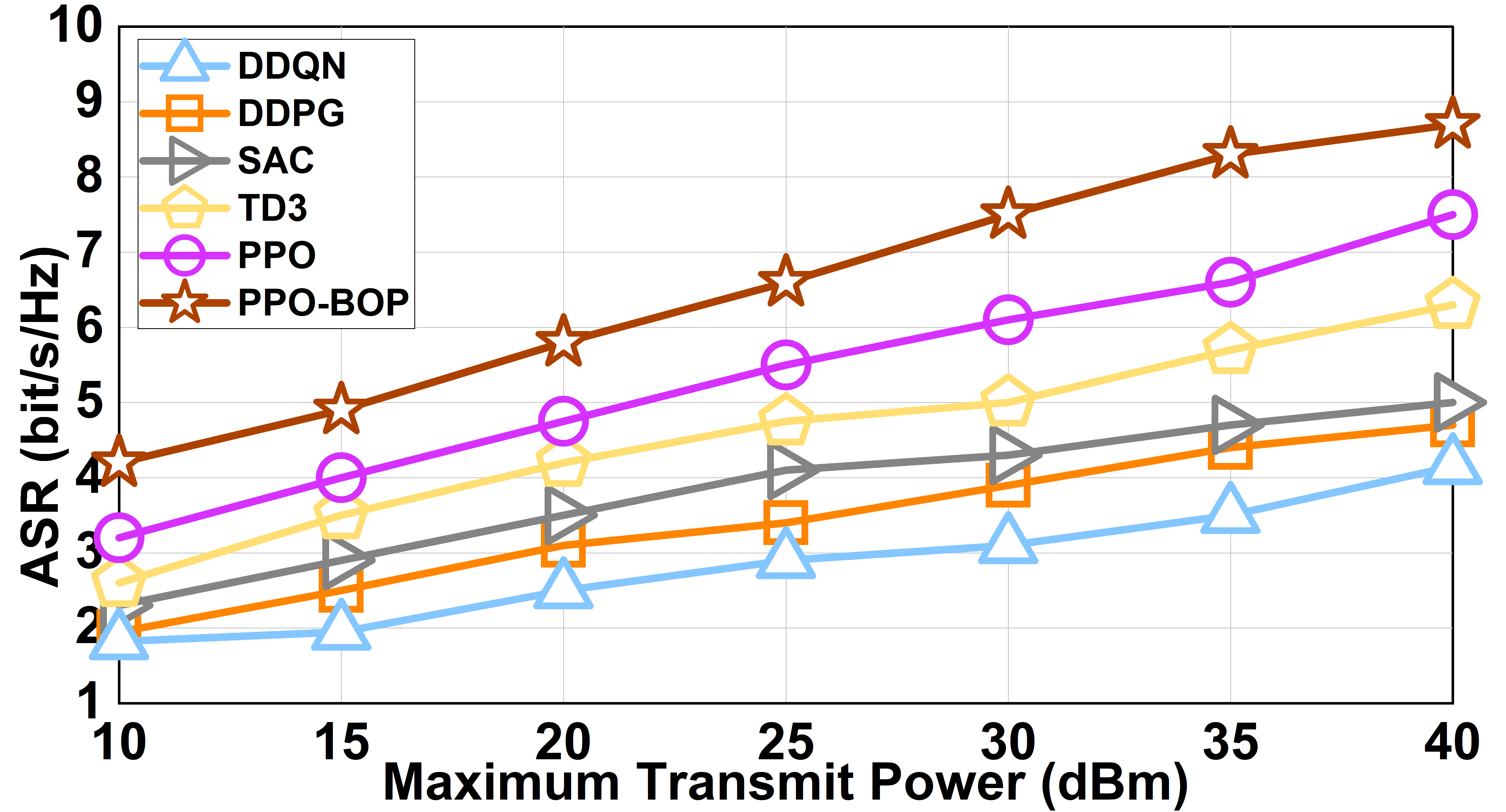}
\caption{Performance comparison of different algorithms in terms of ASR with varying  maximum transmit power ($M=4$, $N=36$, $R^{\min}=0.5$ bit/s/Hz, $\kappa=0.10$, $lr=0.001$, $b=128$, and $l=3$).}
\label{fig5}
\end{figure}

 Fig. \ref{fig5} illustrates the variation of the ASR with the maximum transmit power of the MUs $P_{\max}$. It can be seen that the ASR of the system increases significantly with increasing maximum transmit power $P_{\max}$ for all algorithms. The PPO-BOP exhibits optimal performance with the variation of $P_{\max}$, which demonstrates excellent robustness and stability under varying power conditions.
\subsection{Comparison with Different Parameters}
\par \subsubsection{Impact of Different RHIs} 
\begin{figure}[t]
\centering
\includegraphics[width=0.41\textwidth]{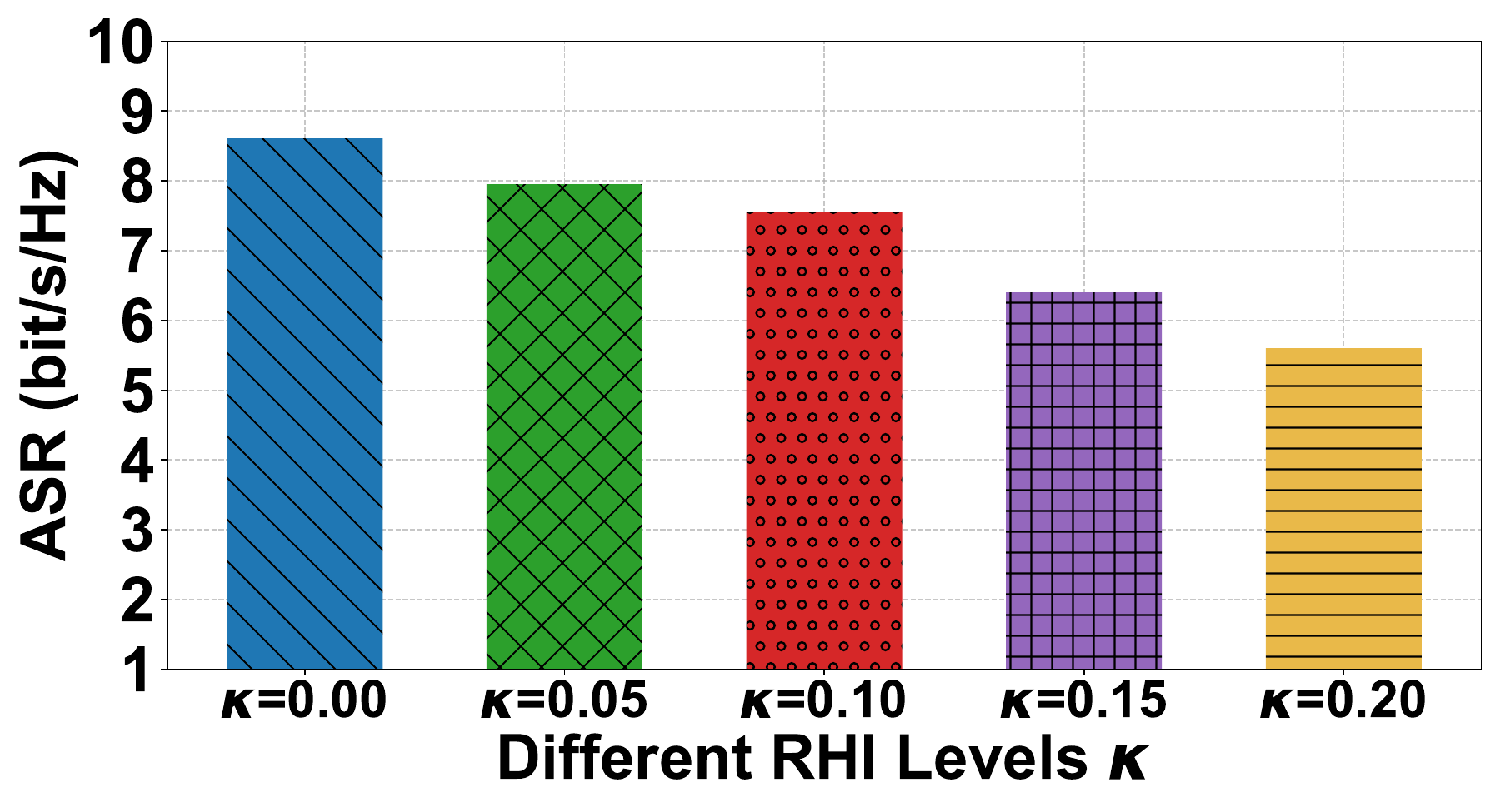}
\caption{The impact of RHI levels on ASR ($M=4$, $N=36$, $P_{\max}=30$ dBm, $lr=0.001$, $b=128$, $l=3$, and $R^{\min}=0.5$ bit/s/Hz).}
\label{fig6}
\end{figure}

Fig. \ref{fig6} illustrates the ASR of the proposed system using PPO-BOP at different RHI levels. As clearly shown, optimal performance is achieved at $\kappa= 0.00$, as RHI interference is minimized under this condition. The ASR decreases significantly as $\kappa$ increases, since RHI introduces distortion noise on the transmitter side, the degrading impact of RHI on the legitimate link being typically more pronounced. Notably, PPO-BOP implicitly learns to counteract these non-linear distortions by finding an optimal power allocation to mitigate RHI effects. This capability confirms the algorithm applicability in practical scenarios with imperfect hardware. This result demonstrates the importance of considering RHI in secure system design.
\par \subsubsection{Impact of Different SIM Parameters} 
\begin{figure}[ht]
\centering
\captionsetup[subfloat]{captionskip=-2pt}
\subfloat[]{\includegraphics[width=1.7in]{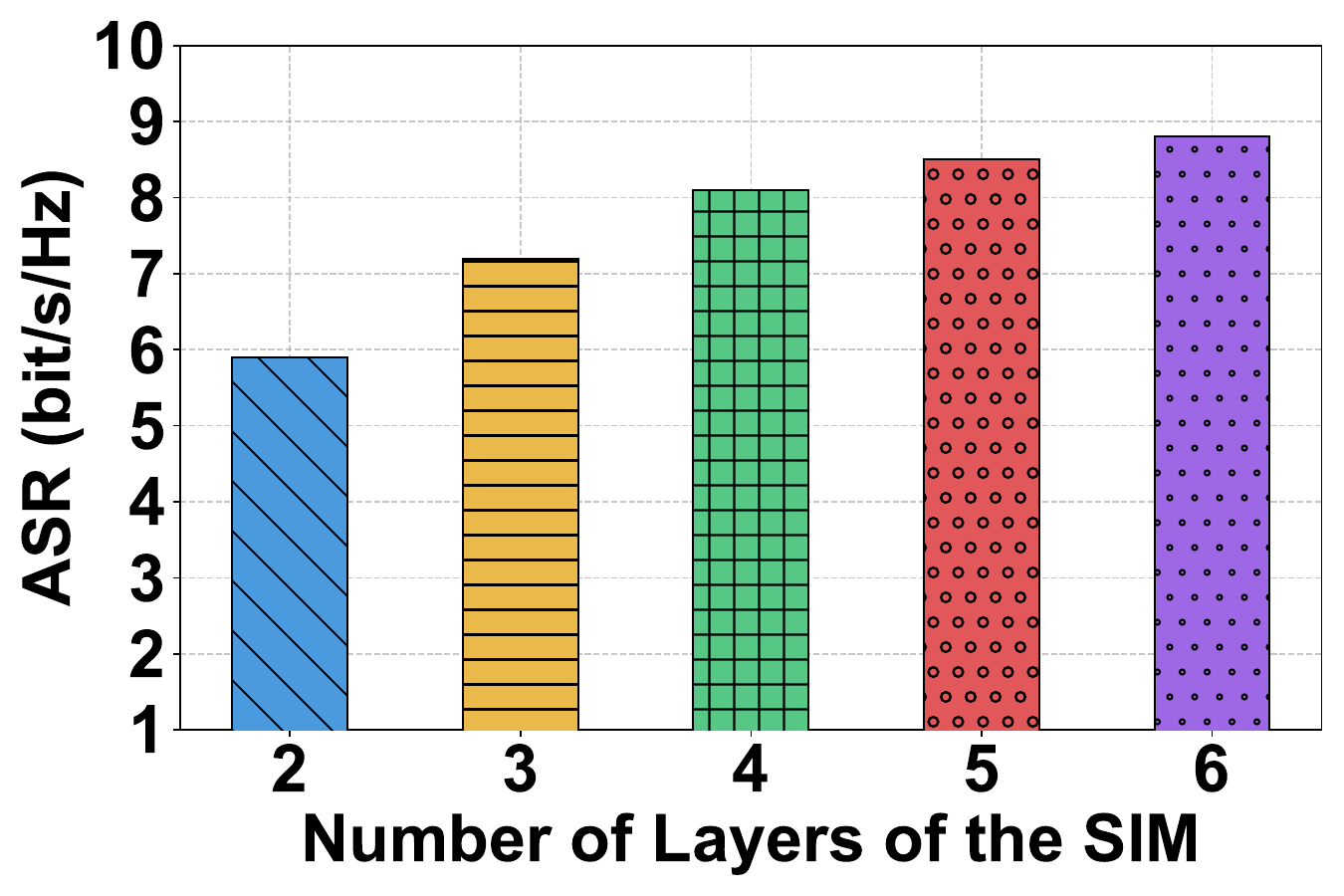}%
\label{fig7_first_case}}
\hfil
\subfloat[]{\includegraphics[width=1.7in]{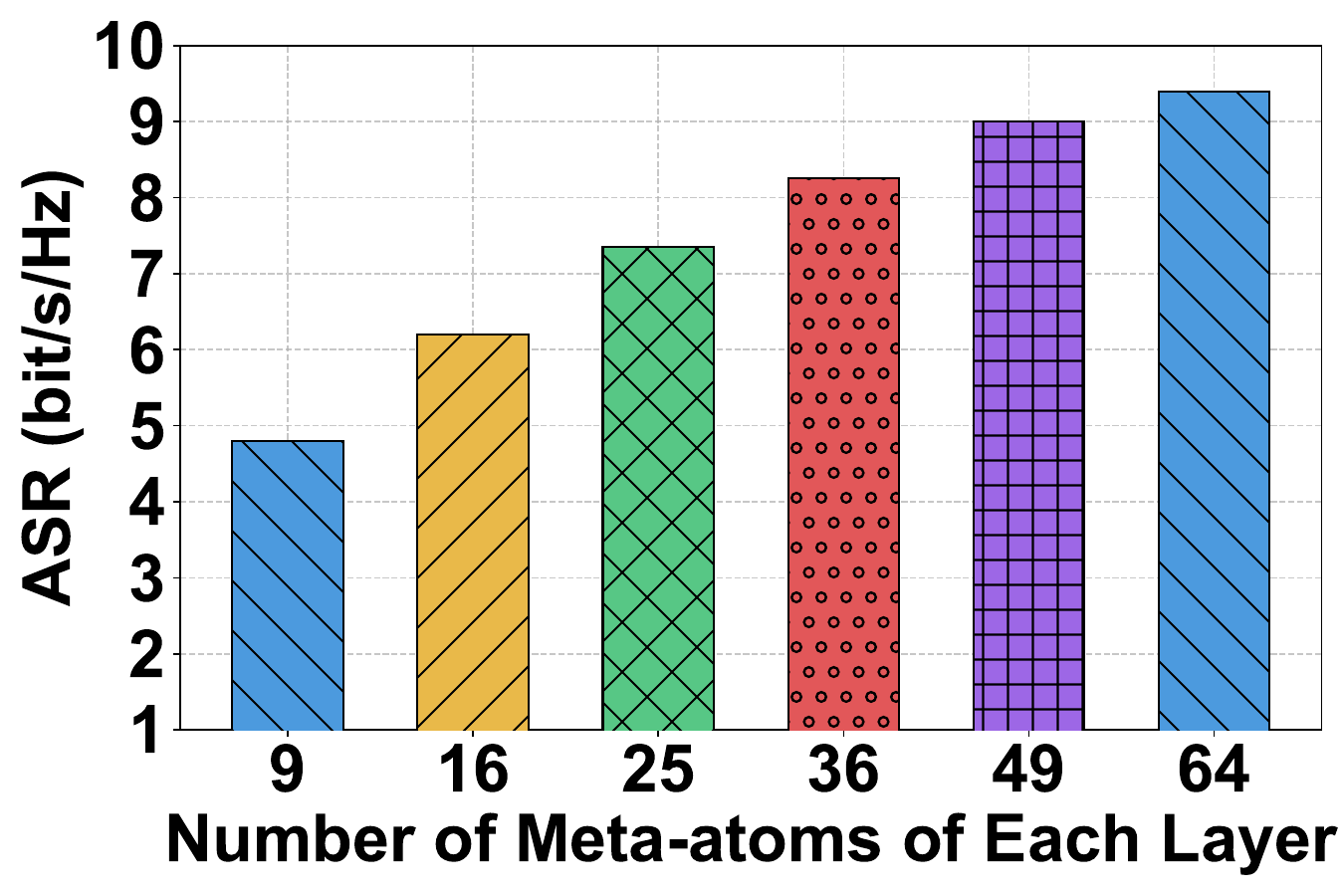}%
\label{fig7_second_case}}
\caption{The impact of SIM parameters on ASR. (a) ASR versus the number of layers. (b) ASR versus the number of meta-atoms of each metasurface layer. ($P_{\max}=30$ dBm, $R^{\min}=0.5$ bit/s/Hz, $\kappa=0.1$, $lr=0.001$, $b=128$, and $l=3$).}
\label{fig_sim_2}
\end{figure}

Fig. \ref{fig_sim_2}(a) shows the ASR versus the number of SIM layers. Specifically, the improvement of ASR is particularly significant when the number of layers increases from 2 to 4. However, once the number of layers exceeds 4, the incremental gain in ASR diminishes, as the overall performance approaches the ceiling imposed by physical or system-level constraints. Fig. \ref{fig_sim_2}(b) illustrates the ASR versus the number of meta-atoms per layer. Similarly, increasing $N$ exhibits diminishing returns in ASR, which is particularly evident when $N$ increases from 49 to 64. This is since the system approaches the maximum secrecy capacity attainable by beamforming, given the specific locations of the MUs and the eavesdropper. This saturation reveals a critical trade-off where further expanding the SIM structure incurs higher hardware manufacturing costs without yielding proportional performance gains. Moreover, an excessive number of meta-atoms significantly increases the dimensionality of the action space, thereby imposing a heavier computational burden on the controller for real-time phase updates.
\par \subsubsection{Impact of Different Enhanced Mechanisms} 
\begin{figure}[htbp]
\centering
\includegraphics[width=0.42\textwidth]{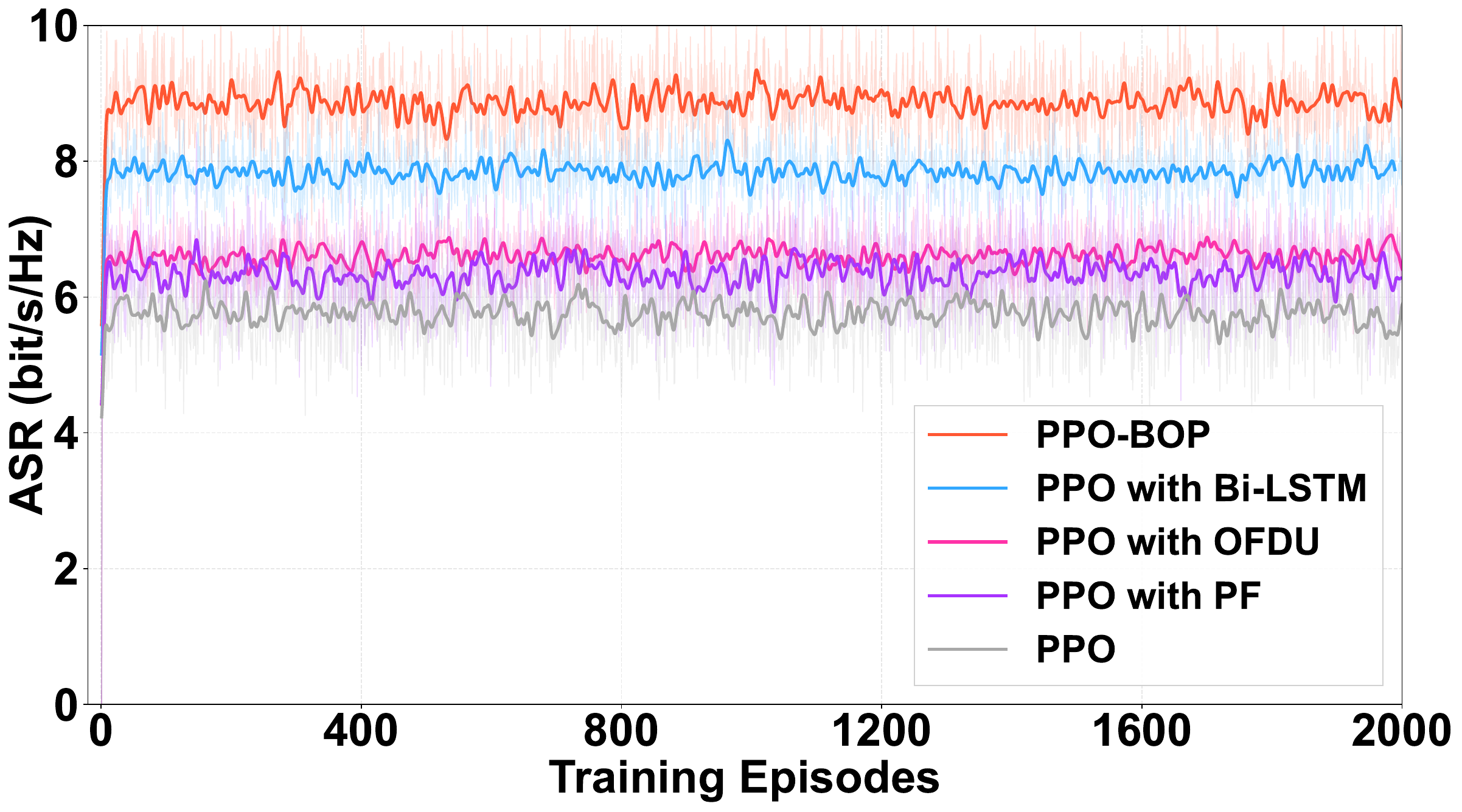}
\caption{Ablation and comparative study of PPO-BOP ($M=4$, $N=36$, $P_{\max}=30$ dBm, $R^{\min}=0.5$ bit/s/Hz, $\kappa=0.1$. $lr=0.001$, $b=128$, and $l=3$).}
\label{fig8}
\end{figure}

We perform an ablation simulation to validate the effectiveness of each enhanced mechanism of the PPO-BOP. As shown in Fig. \ref{fig8}, all algorithms exhibit a growth in rewards during the initial training phase. By integrating enhanced mechanisms, PPO-BOP achieves superior ASR performance, stabilizing at 9.0 to 9.5 bits/s/Hz. The integration of OFDU, PF and Bi-LSTM mechanism effectively guides the PPO-BOP to learn optimal or near-optimal policies, thereby significantly enhancing the performance of ASR.
\par \subsubsection{Impact of Different Network Hyper-parameters} 
\begin{figure}[htbp]
\centering
\captionsetup[subfloat]{captionskip=-1pt}
\subfloat[]{\includegraphics[width=1.7in, height=1.3in]{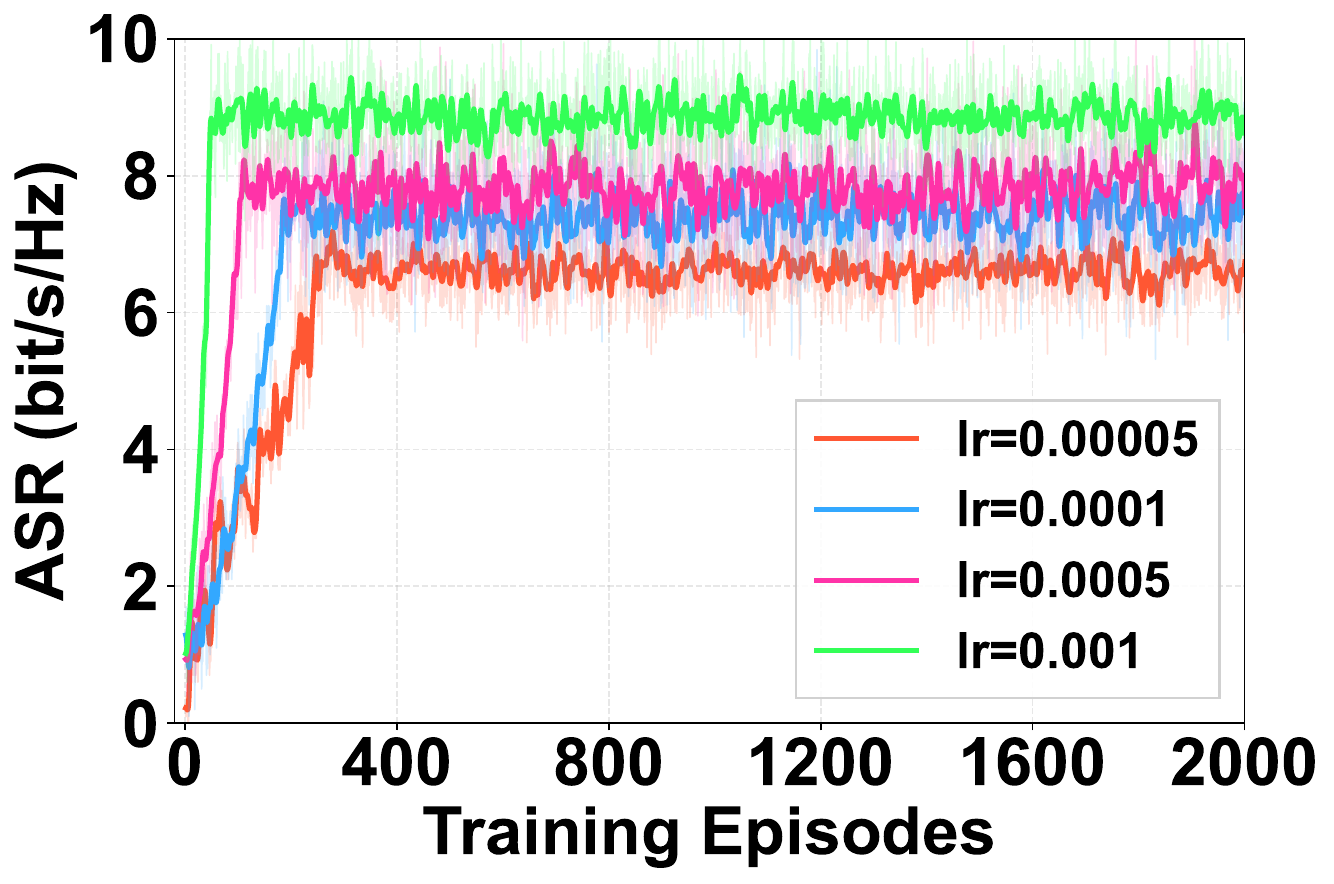}%
\label{fig9_first_case}}
\hfil
\subfloat[]{\includegraphics[width=1.7in, height=1.3in]{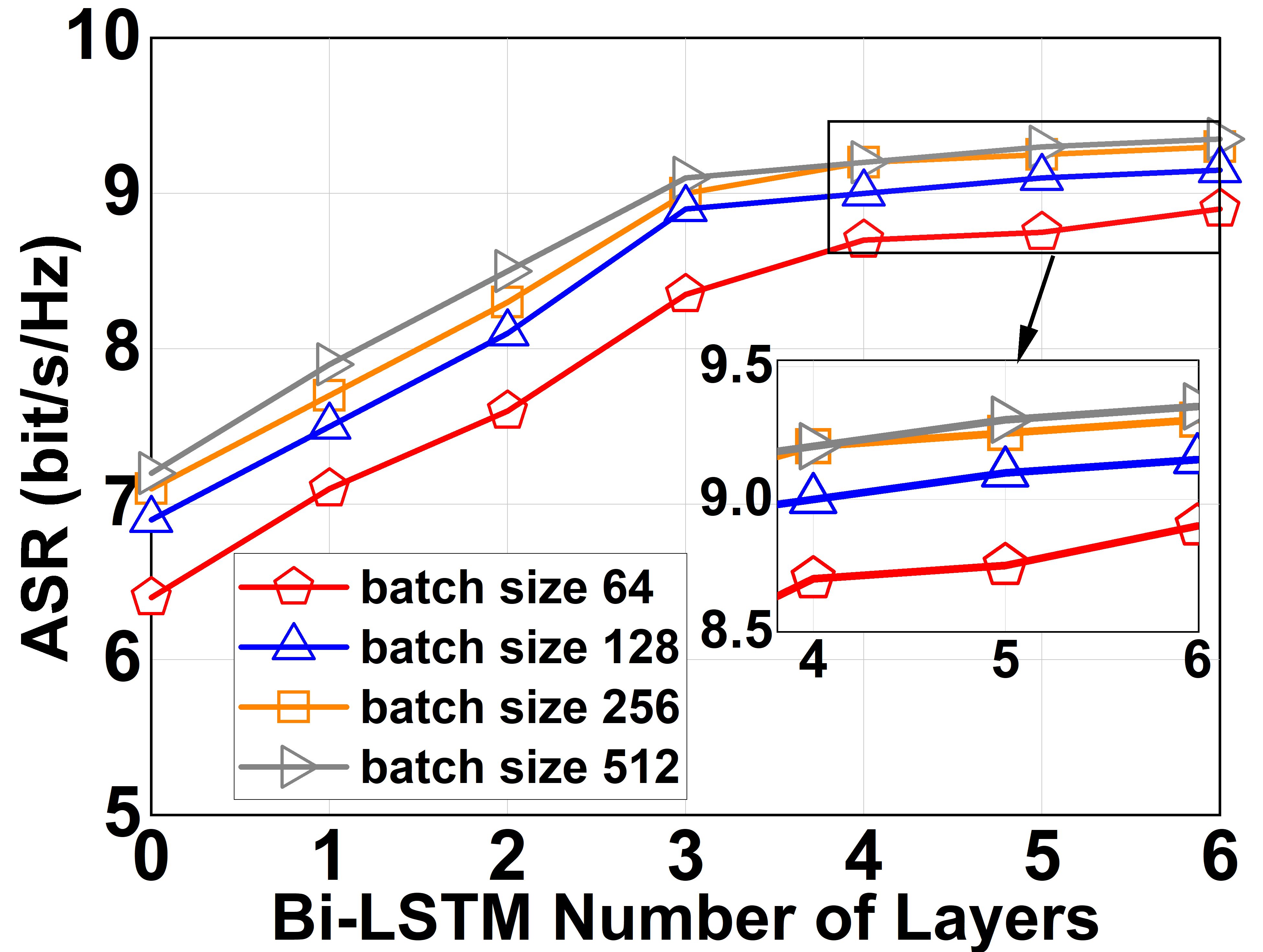}%
\label{fig9_second_case}}
\caption{The impact of PPO-BOP hyper-parameters on ASR. (a) Impact of different $lr$ on ASR. (b) Impact of Bi-LSTM layers and batch sizes on ASR. ($M=4$, $N=36$, $P_{\max}=30$ dBm, $R^{\min}=0.5$ bit/s/Hz, $\kappa=0.1$ $lr=\{0.00005, 0.001, 0.0005, 0.001\}$, and $b=\{64, 128, 256, 512\}$).}
\label{fig_sim3}
\end{figure}

To further enhance the performance of PPO-BOP, we adjust the key hyper-parameters.  Fig. \ref{fig_sim3}(a) illustrates the variation of the ASR over training episodes for different $lr$. As can be seen, the case of $lr=0.001$ achieves the best performance, it achieves the fastest convergence while also securing the highest ASR after convergence. Increasing further $lr$ can lead to too long policy update steps, which can cause instability in the training process. Fig. \ref{fig_sim3}(b) illustrates the impact of varying the number of Bi-LSTM layers and training batch sizes on ASR. When the number of Bi-LSTM layers increases further beyond 3, the growth rate of the ASR is noticeably reduced. This indicates a decrease in marginal returns from increasing network depth, and further increases may result in negligible performance gains. PPO-BOP achieves its highest ASR when the batch size is 512, as this configuration facilitates a smoother convergence to an optimal region of the loss function. However, a larger batch size increases the computational resources required for training and may lead to diminishing returns.

\section{Conclusion}
\label{Conclusion}
\par In this work, we investigated the SIM-assisted secure communication for MUs under eavesdropping threats. For this purpose, we formulated a JPPSOP aimed at maximizing the ASR of the system, which was non-convex. Considering the challenges posed by the MU mobility and channel dynamics, we proposed an enhanced DRL algorithm, termed PPO-BOP. Extensive simulation results validated its effectiveness, showing that PPO-BOP significantly outperformed various strategies and other DRL algorithms in terms of ASR. Furthermore, PPO-BOP maintained robustness and achieved near-optimal performance under different levels of RHI, and different SIM parameters. Ablation studies were conducted to verify that each of the proposed mechanisms played a crucial role in improving algorithm performance. Finally, we analyzed the sensitivity of PPO-BOP to key network hyper-parameters by tuning their values. Future work could extend this framework to more complex eavesdropping scenarios and robust design under imperfect CSI conditions.

\ifCLASSOPTIONcaptionsoff
  \newpage
\fi


\bibliographystyle{ieeetr} 
\bibliography{ref-SIM-Secure}

%




\end{document}